\documentstyle[12pt,epsfig,cite]{article}
\newcommand{\Prob}{\textrm{Pr}}
\newcommand{\beq}{\begin{equation}}
\newcommand{\enq}{\end{equation}}
\newcommand{\beqa}{\begin{eqnarray}}
\newcommand{\enqa}{\end{eqnarray}}

\newcommand{\beql}[1]{\begin{equation}\label{#1}}

\newcommand{\be}{\beta}

\newcommand{\qed}{\hfill $\Box$}

\newcommand{\eq}{\begin{equation}}
\newcommand{\en}{\end{equation}}
\newcommand{\ea}{\begin{eqnarray}}
\newcommand{\an}{\end{eqnarray}}

\newtheorem{theorem}{Theorem}
\newtheorem{lemma}[theorem]{Lemma}
\newtheorem{definition}[theorem]{Definition}

\def\bbC{{\sf C}\kern -6pt {\sf C}}
\def\bbF{{\sf F}\kern -5pt {\sf F}}
\def\bbR{{\sf R}\kern -6pt {\sf R}}
\def\bbZ{{\sf Z}\kern -5pt {\sf Z}}
\def\sfbegin{\begingroup\sf}
\def\sfend{\endgroup}

\def\be{\begin{eqnarray*}}
\def\ee{\end{eqnarray*}}

\begin{document}

\title{On the Throughput-Delay Tradeoff in Cellular Multicast}
\author{Praveen Kumar Gopala and Hesham El Gamal}
\maketitle

\begin{abstract}

In this paper, we adopt a cross layer design approach for
analyzing the throughput-delay tradeoff of the multicast channel
in a single cell system. To illustrate the main ideas, we start
with the single group case, i.e., pure multicast, where a common
information stream is requested by all the users. We consider
three classes of scheduling algorithms with progressively
increasing complexity. The first class strives for minimum
complexity by resorting to a static scheduling strategy along with
memoryless decoding. Our analysis for this class of scheduling
algorithms reveals the existence of a static scheduling policy
that achieves the optimal scaling law of the throughput at the
expense of a delay that increases exponentially with the number of
users. The second scheduling policy resorts to a higher complexity
incremental redundancy encoding/decoding strategy to achieve a
superior throughput-delay tradeoff. The third, and most complex,
scheduling strategy benefits from the cooperation between the
different users to minimize the delay while achieving the optimal
scaling law of the throughput. In particular, the proposed
cooperative multicast strategy is shown to simultaneously achieve
the optimal scaling laws of both throughput and delay. Then, we
generalize our scheduling algorithms to exploit the multi-group
diversity available when different information streams are
requested by different subsets of the user population. Finally, we
discuss the effect of the potential gains of equipping the
base station with multi-transmit antennas and present simulation
results that validate our theoretical claims.

\end{abstract}

\section{Introduction}

Traditional information theoretic investigations pay little, if
any, attention to the notion of delay. Clearly, this approach is
not adequate for many applications, especially those with strict
Quality of Service (QoS) constraints. To avoid this shortcoming,
recent years have witnessed a growing interest in cross layer
design approaches. The underlying idea in these approaches is to
jointly optimize the physical, data link, and networking layers in
order to satisfy the QoS constraints with the minimum expenditure
of network resources. Early investigations on cross layer design
have focused on the single user case \cite{berry,ashu}. These
works have shed light on the fundamental tradeoffs in this
scenario and devised efficient power and rate control policies
that approach these limits. More recent works have considered
multi-user cellular networks \cite{yeh,zhang,berry2,sanjay}. These
studies have enhanced our understanding of the fundamental limits
and the structure of optimal resource allocation strategies. Here,
we take a first step towards generalizing this cross layer
approach to the wireless multicast scenario. This scenario is
characterized by a strong interaction between the network, medium
access, and physical layers. This interaction adds significant
complexity to the problem which motivated the adoption of a
simplified on-off model for the wireless channel in several of the
recent works on wireless multicast \cite{Sasw,Tay,Aceves}. In the
sequel, we argue that employing more accurate models for the
wireless channel allows for valuable opportunities for exploiting
the wireless medium to yield performance gains. More specifically,
our work sheds light on the role of the following characteristics
of the wireless channel in the design of multicast scheduling
strategies: 1) The {\em multi-user diversity} resulting from the
statistically independent channels seen by the different users
\cite{Knopp}, 2) The {\em wireless multicast gain} resulting from
the fact that any information transmitted over the wireless
channel is {\em overheard} by all users, possibly with different
attenuation factors, and 3) The {\em cooperative gain} resulting
from antenna sharing between users \cite{erkip}.

To illustrate the main ideas, we first focus on the single group
(pure multicast) scenario where the same information stream is
transmitted to all users in the network \cite{Prav}. We consider
three classes of scheduling algorithms with progressively
increasing complexity. The first class strives for minimum
complexity by resorting to a static scheduling strategy along with
memoryless decoding\footnote{Memoryless decoding refers to the
fact that the decoder memory is flushed in case of decoding
failure.}. In this approach, we schedule transmission to a
fraction of the users that enjoy favorable channel conditions.
While the identity of the target users change, based on the
channel conditions, the static nature of the algorithm is
manifested in the fact that a {\bf fixed} fraction of the users is
able to decode every transmitted packet. We establish the
throughput-delay tradeoff allowed by varying the fraction of users
targeted in every transmission. To gain more insight into the
problem, we study in more detail the three special cases of
scheduling transmissions to the best, worst and median
user\footnote{These notions will be defined rigorously in the
sequel.}. Here we establish the asymptotic throughput optimality of
the median user scheduler and show that the price for this
optimality is an exponential growth in delay with the number of
users. The second scheduling policy resorts to a higher complexity
incremental redundancy encoding/decoding strategy to achieve a
better throughput-delay tradeoff. This scheme is based on a hybrid
Automatic Repeat reQuest (ARQ) strategy and is shown to yield a
significant reduction in the delay, compared with the median user
scheduler, at the expense of a minimal penalty in the throughput.
The third, and most complex, scheduling strategy benefits from the
cooperation between the different users to minimize the delay
while achieving the optimal scaling law of the throughput. More
specifically, we show that the proposed cooperative multicast
strategy simultaneously achieves the optimal scaling laws of both
throughput and delay at the expense of a high complexity. Finally,
we extend our study to the multi-group scenario where independent
streams of information are transmitted to different groups of
users. Here, we generalize our scheduling algorithms to exploit
the multi-group diversity available in such scenarios.

The rest of the paper is organized as follows. In
Section~\ref{model}, we introduce the system model along with our
notation. In Section~\ref{singleg}, we propose the three classes
of scheduling algorithms for the pure multicast scenario and
characterize the achieved throughput-delay tradeoffs. We then
extend our schemes to exploit the multi-group diversity in
Section~\ref{multig}. The potential performance gains allowed by
multi-transmit antenna base stations are quantified in
Section~\ref{Chifad}. In Section~\ref{sim}, we present numerical
results that validate our theoretical claims in certain
representative scenarios. Finally, some concluding remarks are
offered in Section~\ref{conc}. In order to enhance the flow of the
paper, we collect all the proofs in the Appendices.

\section{System Model} \label{model}

We consider the downlink of a single cell system where a base
station serves $G$ groups of users. The information streams
requested by the different groups from the base station are
independent of each other. Each group consists of $N$ users. All
the users within a group request the same information from the
base station. Unless otherwise stated, the base station is assumed
to be equipped with a single transmit antenna. Each user is
assumed to have only a single receive antenna. We consider
time-slotted transmission in which the signal received by user $i$
in time slot $k$ is given by
\[ y_i[k] = h_i x[k] + n_i[k], \] where $x[k]$ denotes the
complex-valued signal transmitted by the base station in slot $k$,
$h_i$ represents the complex flat fading coeff\mbox{}icient of the
channel between the base station and the $i^{th}$ user, and
$n_i[k]$ represents the zero-mean unit-variance complex additive
white Gaussian noise at the $i^{th}$ user in slot $k$. The noise
processes are assumed to be circularly symmetric and independent
across users. The channel between the base station and each user
is assumed to be quasi-static with coherence time $T_c$. Thus the
fading coeff\mbox{}icients remain constant throughout an interval
of length $T_c$ and change independently from one interval to the
next. The fading coeff\mbox{}icients $\{h_i\}$ are assumed to be
independent and identically distributed (i.i.d.) across the users
and follow a Rayleigh distribution with $E\left[|h_i|^2\right]
= 1$. In this paper, we restrict our attention to this symmetric
scenario, and hence, issues related to fairness are outside the
scope of this work. Each packet transmitted by the base station
is assumed to be of constant size $S$. We further employ the
following short term average power constraint
\[ E\left[ |x[k]|^2 \right] \le P. \]

Clearly, further performance gain may be reaped through a carefully
constructed power allocation policy if this short term power
constraint is replaced by a long term one. This line of work,
however, is not pursued here and we only rely on rate adaptation
and scheduling based on the instantaneous channel state. The
scheduling schemes proposed in the sequel require one further
assumption. We require all the channel gains to be available at
the base station. Hence the proposed scheduling strategies, except
the incremental redundancy scheme\footnote{For the incremental
redundancy scheme, the base station only needs to know when to
stop transmission of the current codeword.}, assume perfect
knowledge of the channel state information (CSI) at both the
transmitter and receiver. In our throughput analysis, we use
capacity expressions for the channel transmission rates. Here we
implicitly assume that the base station employs coding schemes
that approach the channel capacity which justify our use of the
fundamental information theoretic limit of the channel.

In our delay analysis, we consider backlogged queues, and hence,
the only meaningful measure of delay is the transmission delay.
This leads to the following def\mbox{}initions
for throughput and delay that will be adopted in the sequel.

\begin{definition}
The {\em \bf throughput} of a scheduling scheme is defined as the
sum of the throughputs provided by the base station to all the
individual users within all the
groups in the system. \\[0.005in]
\end{definition}

\begin{definition}
The {\em \bf delay} of a scheduling scheme is def\mbox{}ined as
the delay between the instant representing the start of
transmission of a packet belonging to a particular group of users,
and the instant when the packet is successfully decoded by all the
users in that group.
\end{definition}

\vspace{0.005in} A brief comment on the notion of delay adopted in
our work is now in order. This definition suffers from the
fundamental weakness that it does not account for the queuing delay
experienced by the packets. Unfortunately, at the moment we do not
have an analytical characterization of the queuing delay for the
general case. However, as argued in the sequel, our delay analysis
offers a lower bound on the {\em total} delay which is very tight
in several important special cases. Furthermore, this analysis
provides a very useful tool for rank-ordering the different
classes of scheduling algorithms and sheds light on their
structural properties.

To facilitate analytical tractability, we focus on evaluating the
asymptotic scaling laws of the throughput and delay in the sequel.
In this analysis, we use the following set of Knuth's asymptotic
notations throughout the paper: 1)$f(n) = O(g(n))$ iff there are
constants $c$ and $n_0$ such that $f(n) \le c g(n)$ $\forall n >
n_0$, 2) $f(n) = \Omega(g(n))$ iff there are constants $c$ and
$n_0$ such that $f(n) \ge c g(n)$ $\forall n
> n_0$, and 3) $f(n) = \Theta(g(n))$ iff there are constants
$c_1$, $c_2$ and $n_0$ such that $c_1 g(n) \le f(n) \le c_2 g(n)$
$\forall n > n_0$. Furthermore, the two following technical
assumptions are imposed.
\begin{enumerate}
\item We let  \beq \label{Tc} T_c = \Theta \left( \frac{1}{\log
\log NG} \right). \enq This technical assumption is made to ensure
(as shown in the sequel) that the average service time required
for transmitting a packet is not dominated by the scaling behavior
of $T_c$.

\item In our delay analysis, we make an exponential server
assumption, i.e., the rate of service $R$ offered by the server in
any time slot is assumed to follow an exponential distribution
with the same mean as that obtained from our problem formulation.
Thus, for a particular scheduling algorithm, the service rate
distribution is given by
\beq \label{exp}
F_R(r) = 1 - e^{-\mu r}, \quad r \ge 0 ,
\enq
where $\mu = (1/E[R])$ depends on the channel characteristics and
the scheduling algorithm.

\end{enumerate}

\section{Single Group (Pure Multicast) Scenario} \label{singleg}

In this section, we consider the pure multicast scenario where the
same information stream is transmitted to all users in the
network. In the non-cooperative scenario, the throughput-optimal
scheme is an $N$-level superposition coding/successive decoding
scheme \cite{cover}. This strategy, however, suffers from
excessive complexity and the corresponding delay analysis seems
intractable at the moment. This motivates our work where we focus
on the throughput-delay tradeoff of low complexity scheduling
schemes. Interestingly, we identify a low complexity static
scheduling scheme, as defined in the next section, that achieves
the optimal scaling law of the throughput. Furthermore, we
establish the optimality of the proposed cooperative multicast
scheme in terms of the scaling laws of both delay and throughput.

\subsection{Static Scheduling With Memoryless Decoding} \label{singlegen}

In this class of scheduling algorithms, referred to as static
schedulers in the sequel, we schedule transmission to a {\em
fixed} fraction of the users with favorable channel conditions.
The transmission rate is adjusted such that each transmission by
the base station is intended for successful reception by
$(N/\alpha)$ users in the system. Hence at any time instant, the
base station transmits to the user whose instantaneous SNR
occupies the $(N-(N/\alpha)+1)^{th}$ position in the ordered list
of instantaneous SNRs of all users. The other $((N/\alpha)-1)$
users with higher channel gains can also decode the transmitted
information. The parameter $\alpha$ of the scheme is restricted to
be a factor of $N$ and satisfies $\alpha \in {\mathcal Z}^+$ and
$1 \le \alpha \le N$. This scheme is ``static" in the sense that
the fraction of users targeted in every transmission remains the
same (i.e., the parameter $\alpha$ is not a function of time).
When $\alpha> 1$, some of the users will not be able to decode.
The memoryless property dictates that those users {\em flush}
their memories and wait for future re-transmissions of the packet.
This assumption is imposed to limit the complexity of the
encoding/decoding process. In Section~\ref{singleinc}, we relax
this memoryless decoding assumption and quantify the gains offered
by carefully constructed ARQ schemes. As shown later, this class
of static scheduling algorithms exploit both the multi-user
diversity and multicast gains, to varying degrees, depending
on the parameter $\alpha$.

The average throughput of this general static scheduling scheme is given by
\[ R_{tot} = \left( \frac{N}{\alpha} \right) E[R_{\alpha}], \]
where $R_{\alpha}$ is the transmission rate to each of the
intended $(N/\alpha)$ users and is given by \beq \label{servrate}
R_{\alpha} = \log \left(1 + |h_{\pi (N-\frac{N}{\alpha}+1)}|^2 P
\right), \enq where $|h_{\pi (N-\frac{N}{\alpha}+1)}|^2$ is the
channel power gain of the user whose SNR occupies the
$(N-(N/\alpha)+1)^{th}$ position in the ordered list of SNRs of
all users. Throughout the paper, the $\log(.)$ function refers to
the natural logarithm, and hence, the average throughput is
expressed in nats.

\begin{figure}
\centering
\includegraphics[width=\textwidth]{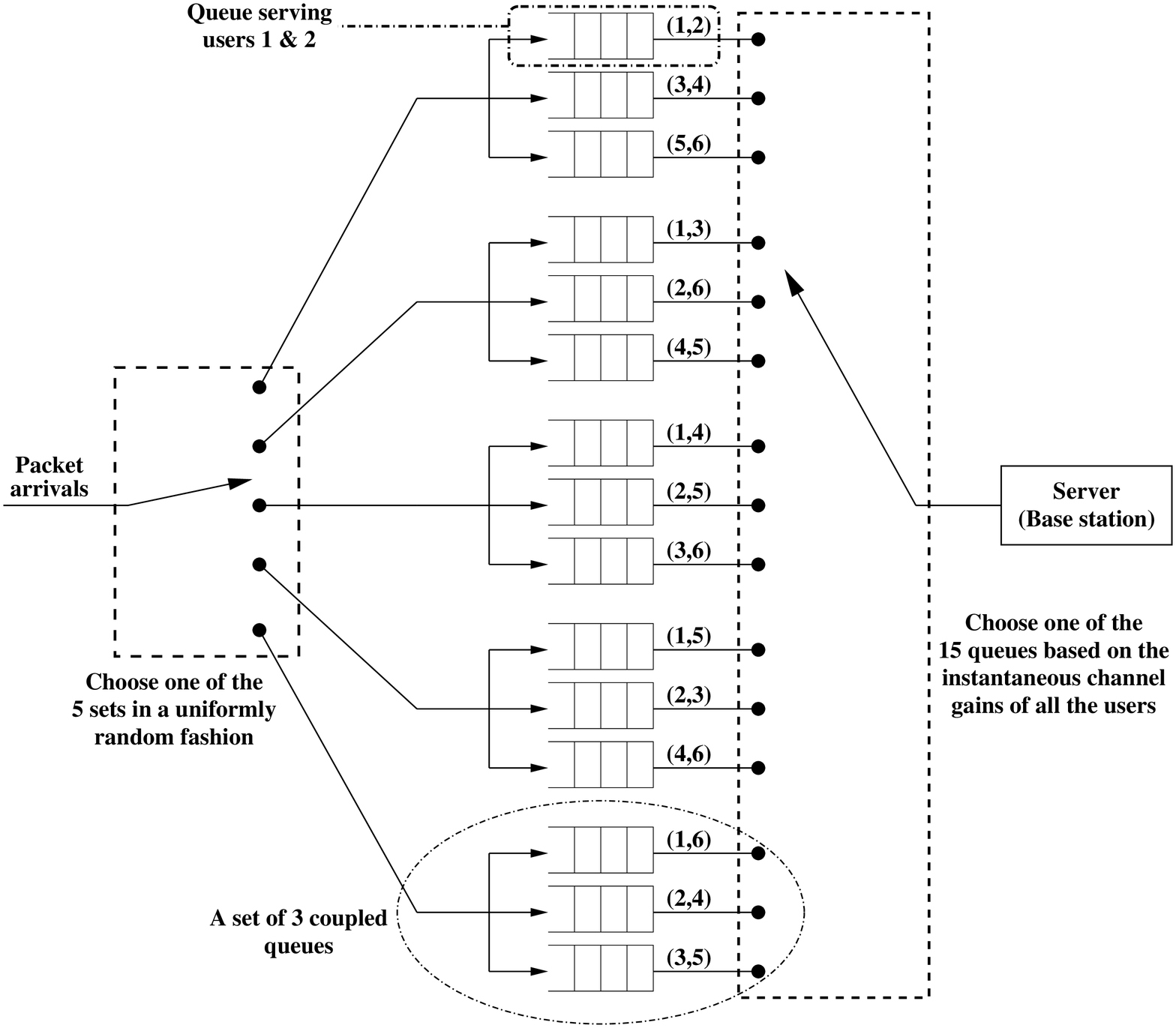}
\caption{A queuing model for a system with $N=6$ users and
$\alpha=3$ \label{quemod}}
\end{figure}

A critical step in the delay analysis is to identify the queuing
model. In our model, the base station maintains ${N \choose
N/\alpha}$ queues, one for each combination of $(N/\alpha)$ users.
These queues can be divided into sets with $\alpha$ {\em coupled}
queues in each set such that the combinations of users served by
the $\alpha$ queues within a set are mutually exclusive (to ensure
that multiple copies of the same packet are not sent to any of the
users) and collectively exhaustive (to ensure that the packet
reaches all the users), i.e., every user in the system is served
by exactly one of the $\alpha$ queues in each set. For example,
with $N=6$ users and $\alpha=3$, we have $15$ queues divided into
$5$ sets with three queues in each set (One possible set of coupled
queues serve users $\{(1,2),(3,4),(5,6)\}$ and another possible
set may serve users $\{(1,4),(2,5),(3,6)\}$. Note that each user
occurs once and only once in each set). Hence, any packet that
arrives at the base station is routed towards one of the 
sets\footnote{Here, we use a probabilistic approach for
choosing the set with a uniform distribution.} where it is stored
in all the $\alpha$ queues within that set (since it needs to be
transmitted to all the users in the system). Thus the delay in
transmitting a particular packet to all the users is given by the
delay in transmitting that packet from each of the $\alpha$
coupled queues in the corresponding set. Moreover, the base
station services only one of the ${N \choose N/\alpha}$ queues at
any time, which is chosen based on the instantaneous fading
coefficients of all the users. An example of the queuing model
for a system with $N=6$ users and $\alpha=3$ is shown in
Fig.~\ref{quemod}.

In our analysis, we benefit from the concept of worst case delay
proposed in \cite{Hass1} for analyzing the delay in unicast
networks. In this work, the authors characterized the worst case
delay by restating their problem as the ``coupon collector
problem'' which has been studied extensively in the mathematics
literature \cite{Newman,Feller,Johnson}. In the coupon collector
problem, the users are assumed to have coupons and the transmitter
is the collector that selects one of the users randomly (with
uniform distribution) and collects his coupon. The problem is to
characterize the average number of trials required to ensure that
the collector collects $m$ coupons from all the users. Our queuing
problem is analogous to the coupon collector problem with the only
{\em fundamental} difference being that the size of the coupons is
time-varying in our problem due to rate adaptation (the detailed
analysis is presented in the proofs). Now, we are ready to state
our result that characterizes the scaling laws of throughput and
delay for the different static scheduling algorithms.

\begin{theorem} \label{gentd}
The average throughput $R_{tot}$ of the general static scheduling
scheme is given by \beq \label{genthru} R_{tot} = \frac{N}{\alpha}
\left[ \sum_{i=1}^N {N \choose i} (-1)^i
e^{\left(\frac{i}{P}\right)} Ei\left(-\frac{i}{P}\right) +
\sum_{k=(N-\frac{N} {\alpha}+1)}^{N-1} {N \choose k} \left[
\sum_{i=0}^k {k \choose i} (-1)^i e^{\left(\frac{N-k+i}{P}\right)}
Ei\left(-\frac{(N-k+i)}{P}\right) \right] \right], \enq where \[
Ei(x) = \int_{-\infty}^{x} \frac{e^t}{t} {\mathrm d} t. \] The average
delay $D$ of this scheme satisfies \beq \label{gendel} D= \max
\left\{ \Omega \left( {N \choose N/\alpha} \frac{\log \alpha}{\log
\log N} \right), \Omega \left( {N \choose N/\alpha} E[X_{min}]
\right) \right\}, \enq where $X_{min}=\min_{i=1}^{\alpha} X_i$ and
the $X_i$'s are defined as the service times required for
transmitting a packet from the $i^{th}$ queue of a set of $\alpha$
queues assuming that the server always services the $i^{th}$
queue.
\end{theorem}
\vspace{0.005in}

To gain more insights into the rather involved throughput and
delay expressions of Theorem~\ref{gentd}, we study three special
cases of the general static scheduling scheme in more detail. This
detailed analysis sheds light on the throughput-delay tradeoff
achievable by varying $\alpha$. We further establish the
optimality of the scheduler corresponding to $\alpha=2$ with
respect to the throughput scaling law.

\subsubsection{Worst User Scheduler}

The worst user scheme corresponds to the case $\alpha=1$ of the
general scheduling scheme. This scheme maximally exploits the
multicast gain by always transmitting to the user with the least
instantaneous SNR. This enables all the users to successfully
decode the transmission and thus any particular packet reaches all
the users in a single transmission. However, the multi-user
diversity inherent in the system works against the performance of
this scheme and results in a decrease in the individual throughput
to any user.

The average throughput of the worst user scheme is given by
\[ R_{tot} = N E \left[ \log \left(1 + |h_{\pi (1)}|^2 P\right) \right], \]
where $|h_{\pi (1)}|^2$ is the minimum channel gain among all the $N$ users in the
system, whose distribution and density functions are given by
\[ F_{|h_{\pi (1)}|^2}(x) = 1-e^{-Nx} \quad \textrm{and} \quad f_{|h_{\pi (1)}|^2}(x)
 = Ne^{-Nx}, \quad x\ge 0. \]
For implementing this scheme, the base station needs to maintain only a single
queue that caters to all the users in the system.

\begin{lemma} \label{corw}
The average throughput of the worst user scheme scales as
\beq
R_{tot} = \Theta (1)
\enq
with the number of users $N$. The average delay of this scheme scales as
\beq \label{mindel}
D = \Theta(N).
\enq
\end{lemma}
\vspace{0.005in}

Thus the average throughput of the worst user scheme does not
scale with the number of users $N$ while the average delay
increases linearly with $N$.

\subsubsection{Best User Scheduler}

This scheme corresponds to the case $\alpha=N$ of the general
scheduling scheme and maximally exploits the multi-user diversity
available in the system. Since the transmission rate is adjusted
based on the user with the maximum instantaneous SNR, this scheme
fails to exploit any of the multicast gain and any particular
packet must be {\em repeated} $N$ times. The average throughput of
the best user scheme is given by
\[ R_{tot} =  E \left[ \log \left(1 + |h_{\pi (N)}|^2 P\right) \right], \]
where $|h_{\pi (N)}|^2$ is the maximum channel gain among all the $N$ users in
the system, whose distribution function is given by
\[ F_{|h_{\pi (N)}|^2}(x) = \left(1-e^{-x}\right)^N, \quad x \ge 0. \]
In this special case, the base station maintains $N$ queues, one
for each user in the system, and any packet that arrives into the
system enters all the $N$ queues. The following result establishes
the throughput and delay scaling laws achieved by the best user scheduler.

\begin{lemma} \label{corb}
The average throughput of the best user scheme scales as
\beq
R_{tot} = \Theta ( \log \log N)
\enq
with the number of users $N$. The average delay of this scheme scales as
\beq \label{maxdel}
D = \Omega \left( \frac{N \log N}{\log \log N} \right).
\enq
\end{lemma}
\vspace{0.005in}

From Lemmas~\ref{corw}~and~\ref{corb}, one can conclude that {\em
maximally} exploiting the multi-user diversity yields higher
throughput gains than {\em maximally} exploiting the multicast
gain. This throughput gain, however, is obtained at the expense of
a higher delay. This observation motivates the investigation of
other variants of the static scheduling strategy which achieve
other points on the throughput-delay tradeoff.

\subsubsection{Median User Scheduler}

The median user scheduler corresponds to the case $\alpha=2$ of
the general scheduling scheme. In this scheme, the base station
maintains ${N \choose N/2}$ queues, one for each combination of
$(N/2)$ users. This scheme strikes a balance between exploiting
multi-user diversity and multicast gain. The base station always
transmits to the user whose instantaneous SNR occupies the median
position of the ordered list of SNRs. Each transmission is,
therefore, successfully decoded by half the users in the system
and the same information needs to be repeated only twice before it
reaches all the users. Thus, unlike the best user scheduler, this
scheduler benefits from the wireless multicast gain. Moreover,
unlike the worst user scheduler, the inherent multi-user diversity
does not degrade the performance of this scheduler (since the
instantaneous SNR of the median user is not expected to degrade
with $N$). In fact, we show in the following that this scheme
achieves the optimal scaling law of the throughput as the number
of users $N$ grows to infinity.

\begin{lemma} \label{corm}
The proposed median user scheme achieves the optimal scaling law
of the throughput. The average throughput of this scheme scales as
\beq \label{medscal}
R_{tot} = \Theta(N)
\enq
with the number of users $N$. The average delay of this scheme scales as
\beq \label{meddel}
D = \Theta \left( {N \choose N/2} \right) = \Theta \left( \frac{2^N}{\sqrt{N}} \right).
\enq
\end{lemma}
\vspace{0.005in}

Thus the throughput optimality of the median user scheduler is
obtained at the expense of an exponentially increasing delay with
the number of users $N$. Overall, these three special cases of the
static scheduling strategy show that one can achieve different
points on the throughput-delay tradeoff by varying $\alpha$.

\subsection{Incremental Redundancy Multicast} \label{singleinc}

In this section, we relax the memoryless decoding requirement and
propose a scheme that employs a higher complexity incremental
redundancy encoding/decoding strategy to achieve a better
throughput-delay tradeoff than the static scheduling schemes. The
proposed scheme is an extension of the incremental redundancy
scheme given by Caire {\em et al} in \cite{Caire}. An information
sequence of $b$ bits is encoded into a codeword of length $LM$,
where $M$ refers to the rate constraint. The f\mbox{}irst $L$ bits
of the codeword are transmitted in the f\mbox{}irst attempt. If a
user is unable to successfully decode the transmission, it sends
back an ARQ request to the base station. If the base station
receives an ARQ request from any of the users, it transmits the
next $L$ bits of the same codeword in the next attempt. This
process continues until either all $N$ users successfully decode
the information or the rate constraint $M$ is violated. Then the
codeword corresponding to the next $b$ information bits is
transmitted in the same fashion. In this scheme, even if some of
the users successfully decode the information in very few
attempts, they still have to wait until all the $N$ users
successfully receive the information before any new information is
transmitted to them by the base station. This sub-optimality of
the proposed scheme results in significant complexity reduction by
avoiding the use of superposition coding and successive decoding.
Moreover, this scheme does not require the knowledge of perfect
CSI at the base station. The base station only needs to know when
to stop transmission of the current codeword. Hence the feedback
required is minimal. The following result establishes the superior
throughput-delay tradeoff achieved by this scheme, compared with
the class of static schedulers with memoryless decoding.

\begin{theorem} \label{thinc}
The average throughput of the incremental redundancy scheme scales as
\beq \label{Rinc}
R_{tot} = \Theta \left( \frac{N \log \log N}{\log N} \right)
\enq
with the number of users $N$. The average delay $D$ of this scheme scales as
\beq \label{Dinc}
D = \Theta \left( \frac{\log N}{\log \log N} \right).
\enq
\end{theorem}
\vspace{0.2in}

Thus, we can see that incremental redundancy multicast avoids the
exponentially growing delay of the median user scheduler at the
expense of a minimal penalty in throughput. In fact, the loss in
both delay and throughput scaling laws, compared to the optimal
values, is only a factor of $\log(N)/\log\log(N)$. In this
approach, the base station needs to maintain only a single queue
that serves all the users in the system. This approach, however,
entails added complexity in the incremental redundancy encoding
and the storage and joint decoding of all the observations.

\subsection{Cooperative Multicast} \label{singlecoop}

In this section, we demonstrate the benefits of user cooperation
and quantify the tremendous gains that can be achieved by allowing
the users to cooperate with each other. In particular, we propose
a cooperation scheme that minimizes the delay while achieving the
optimal scaling law of the throughput. This scheme is divided into
two stages. In the first half of each time slot, the base station
transmits the packet to one half of the users in the system (i.e.,
the median user scheduler). During the next half of the slot, the
base station remains silent. Meanwhile all the users that
successfully decoded the packet in the first half of the slot
cooperate with each other and transmit the packet to the other
$(N/2)$ users in the system. This is equivalent to a transmission
from a transmitter equipped with $(N/2)$ transmit antennas to the
worst user in a group of $(N/2)$ users. If $R_{s1}$ and $R_{s2}$
are the rates supported in the first and second stage
respectively, then the actual transmission rate is chosen to be
$\min \{ R_{s1} , R_{s2} \}$ in both stages of the cooperation
scheme. Note that the rate $R_{s2}$ is chosen such that the information
can be successfully decoded even by the worst of the remaining
$(N/2)$ users. Here, we note that this scheme requires the
base station to know the CSI of the inter-user channels. The
scheme, however, does not require the users to have such
transmitter CSI (i.e., in the second stage the users cooperate
blindly by using i.i.d. random coding). The average throughput of
the proposed cooperation scheme is thus given by
\[ R_{tot} = \left( \frac{N}{2} \right) E \left[ \min \{ R_{s1} , R_{s2} \} \right]. \]

The following result establishes the optimality of the proposed
scheme, in terms of both delay and throughput scaling laws.

\begin{theorem} \label{coop}
The proposed cooperation scheme achieves the optimal scaling laws
of both delay and throughput. In particular, the average
throughput of this scheme scales as \beq R_{tot} = \Theta(N) \enq
with the number of users $N$, while the average delay scales as
\beq D = \Theta(1). \enq Here we assume that the inter-user
channels have the same fading statistics as the channels between
the base station and users, and the {\bf total} transmitted power
is upper bounded by $P$.
\end{theorem}
\vspace{0.005in}

The price for this optimal performance is the added complexity
needed to 1) equip every user terminal with a transmitter, 2)
decode/re-encode the information at each cooperating user
terminal, and 3) inform the base station with perfect CSI of the
inter-user channels.

\section{Multi-group Diversity} \label{multig}

In this section, we generalize the scheduling schemes proposed in
Section~\ref{singleg} to the multi-group scenario where different
information streams are requested by different subsets of the user
population. We modify the proposed schemes to exploit the
multi-group diversity available in this scenario by always
transmitting to the best group. We characterize the asymptotic
scaling laws of the throughput and delay of the static
schedulers with the number of users per group $N$ and the number
of groups $G$ in the following theorem.

\begin{theorem} \label{statmg}
\begin{enumerate}
\item{The average throughput of the best among worst users scheme
scales as \beq R_{tot} =  \Theta(\log G) \enq with $N$ and $G$.
The average delay of this scheme scales as \beq \label{mindelmg} D
= \Theta \left( \frac{NG}{\log G} \right). \enq} \item{The average
throughput of the best among best users scheme scales as \beq
R_{tot} = \Theta(\log \log NG) \enq with $N$ and $G$. The average
delay of this scheme scales as \beq \label{maxdelmg} D = \Omega
\left( \frac{N G \log N}{\log \log NG} \right). \enq} \item{The
average throughput of the best among median users scheme satisfies
\beq \label{Rmed} \Omega (N) = R_{tot} = O( N \log \log G) , \enq
while the average delay of this scheme satisf\mbox{}ies \beq
\label{meddelmg} \Omega \left( \frac{G 2^N}{\sqrt{N} \log \log G}
\right) = D = O \left( \frac{G 2^N}{\sqrt{N}} \right). \enq}
\end{enumerate}
\end{theorem}
\vspace{0.005in}

In the multi-group incremental redundancy scheme, the information
bits corresponding to each of the groups are encoded
independently. During each time slot, the base station selects
that group for which it can send the highest total instantaneous
rate to the users who failed to decode up to this point. This
selection process makes the scheme ``dynamic" in the sense that
the outcome of the scheduling process at any particular time slot
depends on the outcomes in all previous slots. Unfortunately, this
dynamic nature of the proposed scheme adds significant complexity
to the problem and, at the moment, we do not have an analytical
characterization of the corresponding scaling laws.

In the multi-group cooperation scheme, during each time slot, the
base station selects the best group $\hat{g}$ for transmission
according to the condition
\beq \label{cond5}
\hat{g} = \arg \max_{g=1,\cdots,G} \left\{ \left( \frac{N}{2} \right)
\min \{ R_{s1}^g , R_{s2}^g \} \right\}.
\enq

\begin{theorem} \label{coopmg}
The average throughput of the proposed multi-group cooperation
scheme satisfies \beq \Omega (N) = R_{tot} = O \left( N \log \log
G \right), \enq while the average delay of this scheme satisfies
\beq \Omega \left( \frac{G}{\log \log G} \right) = D = O (G). \enq
\end{theorem}
\vspace{0.005in}

As expected, the throughput gain resulting from the multi-group
diversity entails a corresponding price in the increased delay.

\section{Multi-Transmit Antenna Gain} \label{Chifad}

The performance of the proposed static scheduling schemes depends
on the spread of the fading distribution. For exploiting
significant multi-user diversity gains, the distribution needs to
be well-spread out. The lower the spread of the distribution, the
lesser the multi-user diversity gain (or loss as shown in the
following). To illustrate this point, we consider a scenario where
the base station is equipped with $L$ transmit antennas. We assume
that the base station has knowledge of only the total effective
SNR at any particular user and does not know the individual
channel gains from each transmit antenna to that user. Under this
assumption, the base station just distributes the available power
equally among all the $L$ transmit antennas. Thus the effective
fading power gains follow a normalized Chi-square distribution
with $2L$ degrees of freedom. Note that the fading power gains are
exponentially distributed (Chi-square with 2 degrees of freedom)
in the single transmit antenna case. We now characterize the
asymptotic scaling laws of the throughput of the proposed static
schedulers for this multi-transmit antenna scenario. Note that all
the results in this section are derived for the case where $L$ is
a constant and does not scale with $N$.

\subsection{Worst User Scheduler}

For the worst user scheme, the average throughput is given by
\[ R_{tot} = N E \left[ \log \left( 1 + |\chi_{min}|^2 P \right) \right], \]
where $|\chi_{min}|^2 = \min_{i=1}^N |\chi_i|^2$, and $|\chi_i|^2$ corresponds
to the effective fading power gain at the $i^{th}$ user that follows a
normalized Chi-square distribution with $2L$ degrees of freedom and
whose distribution function is given by 
\beq \label{chinorm}
F(x) = 1 - e^{-Lx} \left( \sum_{k=0}^{L-1} \frac{(Lx)^k}{k!} \right),
\quad x \ge 0.
\enq

\begin{lemma} \label{Chiw}
When the base station is equipped with $L$ transmit antennas, the
average throughput of the worst user scheme scales as
\beq \label{tchiw}
R_{tot} = \Theta \left( N^{\left(\frac{L-1}{L} \right)} \right).
\enq
\end{lemma}
\vspace{0.005in}

Thus the average throughput increases with $L$. This is expected
since the performance of the worst user scheduler is {\em degraded} by
the tail of the fading distribution. Hence, as $L$ increases, the
spread of the fading distribution decreases, and consequently, the
inherent multi-user diversity has a reduced effect on the
performance of the scheduler. This leads to a rise in the average
throughput of the worst user scheme from $\Theta(1)$ for the
single transmit antenna case to $\Theta(N)$ for large values of
$L$.

\subsection{Best User Scheduler}

For the best user scheme, the average throughput is given by
\[ R_{tot} = E \left[ \log \left( 1 + |\chi_{max}|^2 P \right) \right], \]
where $|\chi_{max}|^2 = \max_{i=1}^N |\chi_i|^2$.

\begin{lemma} \label{Chib}
When the base station is equipped with $L$ transmit antennas, the
average throughput of the best user scheme scales as
\beq \label{tchib}
R_{tot} = \Theta \left( \log  \left( 1 +  \frac{\log N + (L-1) \log
\log N}{L} \right) \right).
\enq
\end{lemma}
\vspace{0.005in}

Since the best user scheduler leverages multi-user diversity to
enhance the throughput, one can see that the throughput of the
best user scheme decreases as $L$ increases.

\section{Numerical Results} \label{sim}

\begin{figure}
\centering
\includegraphics[width=\textwidth]{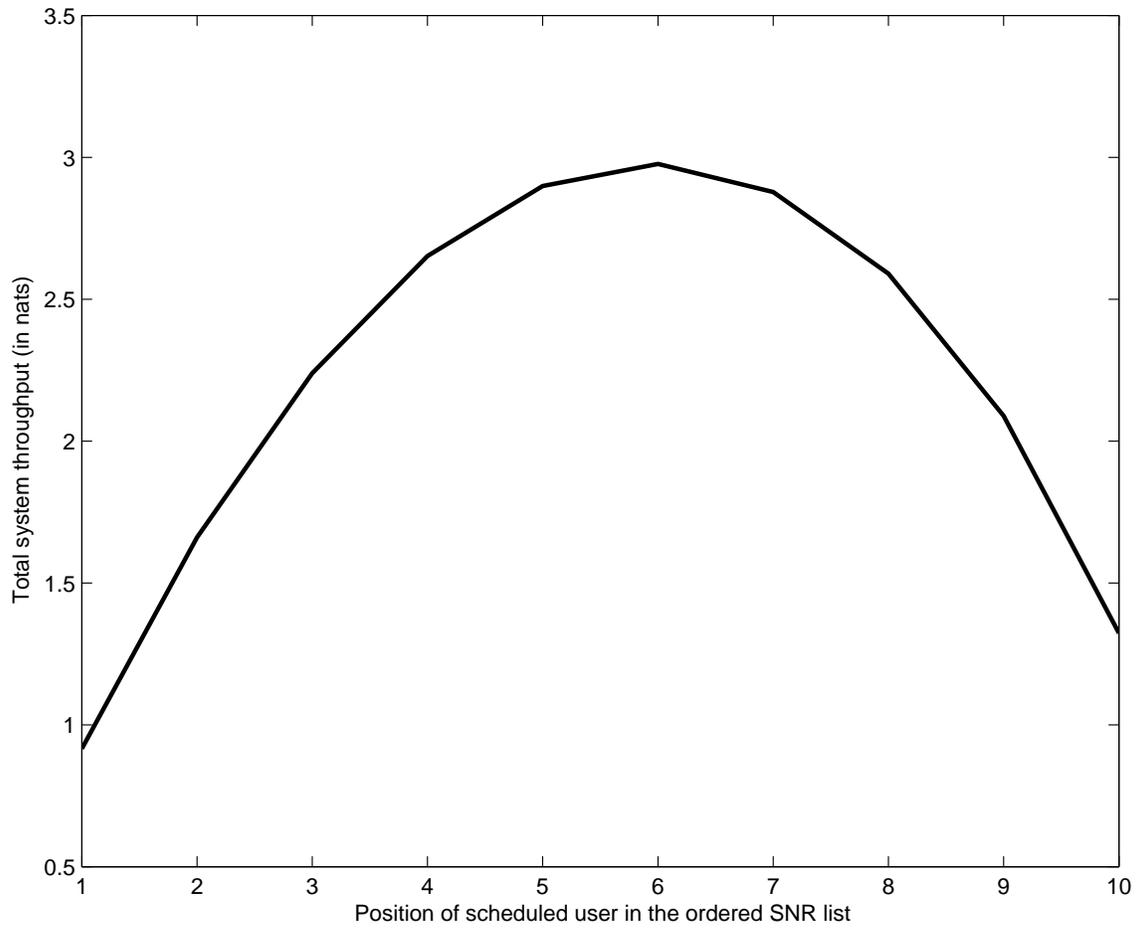}
\caption{Throughput of the general static scheduling scheme for
different positions of the intended user in the ordered list of
SNRs of all users (N=10) \label{tpos}}
\end{figure}

\begin{figure}
\centering
\includegraphics[width=\textwidth]{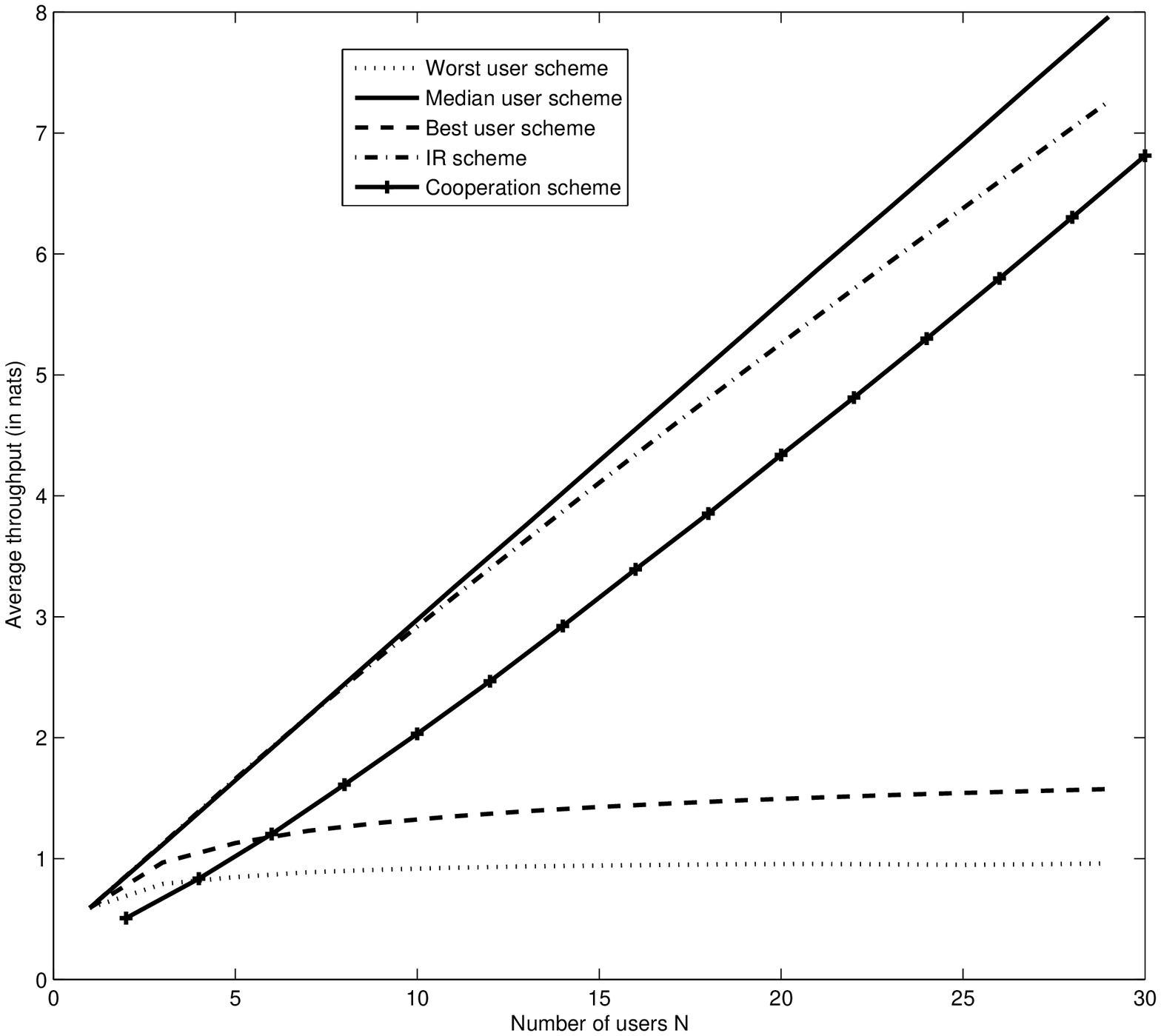}
\caption{Comparison of the throughput of the proposed schemes
\label{compt}}
\end{figure}

\begin{figure}
\centering
\includegraphics[width=\textwidth]{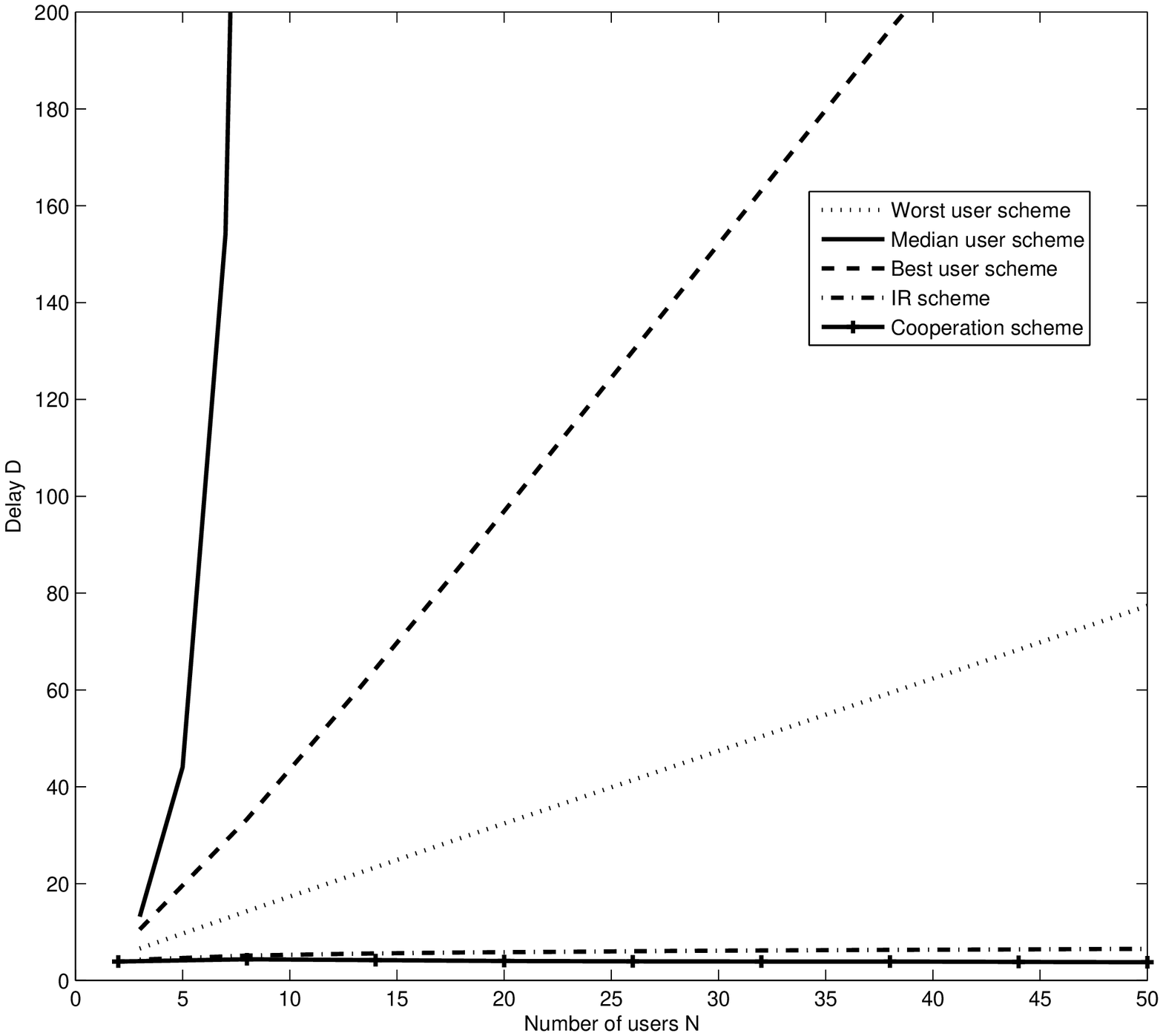}
\caption{Comparison of the delay of the proposed schemes
\label{compd}}
\end{figure}

\begin{figure}
\centering
\includegraphics[width=\textwidth]{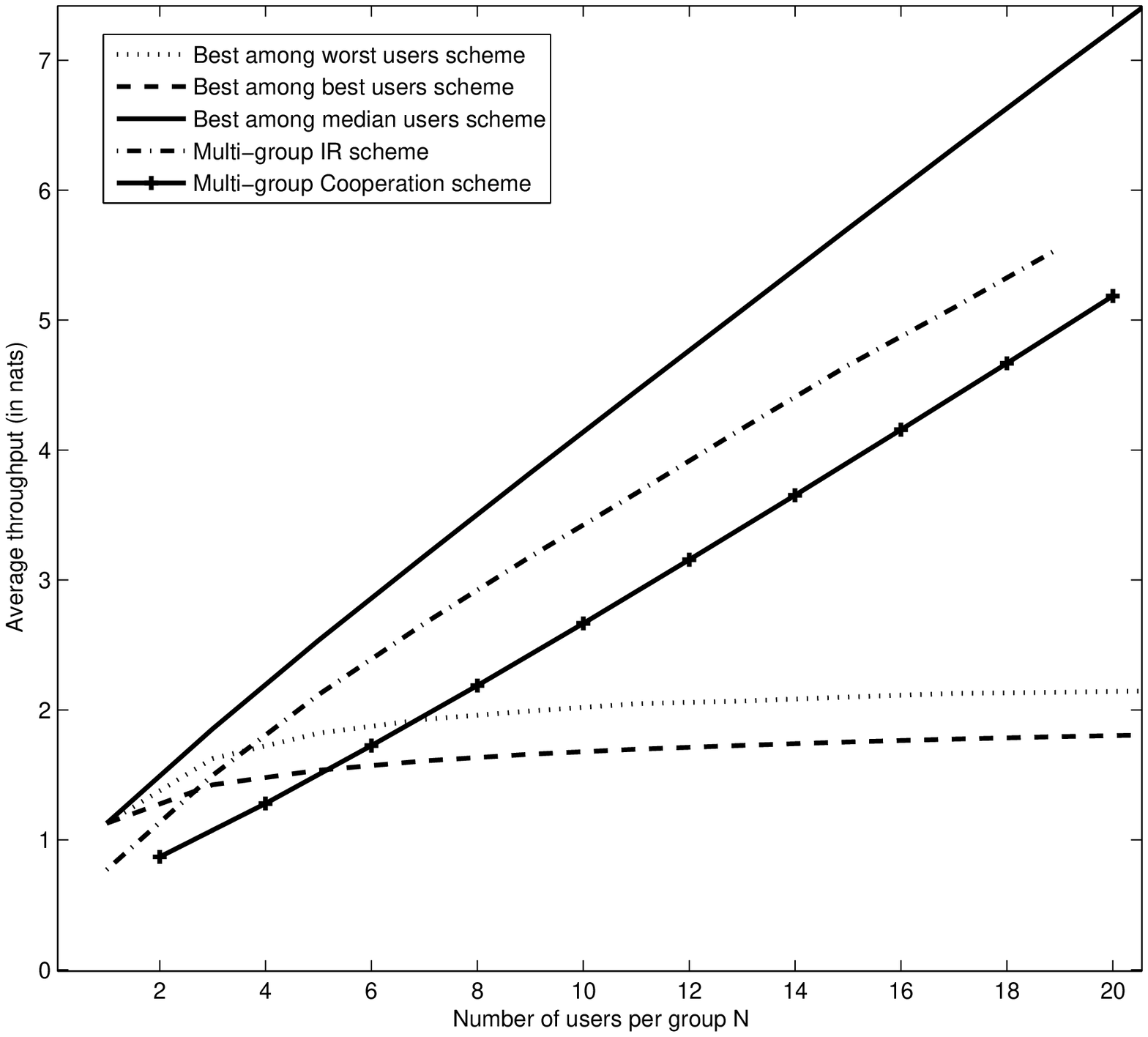}
\caption{Comparison of the throughput of the proposed schemes for
$G=5$ groups \label{t5}}
\end{figure}

\begin{figure}
\centering
\includegraphics[width=\textwidth]{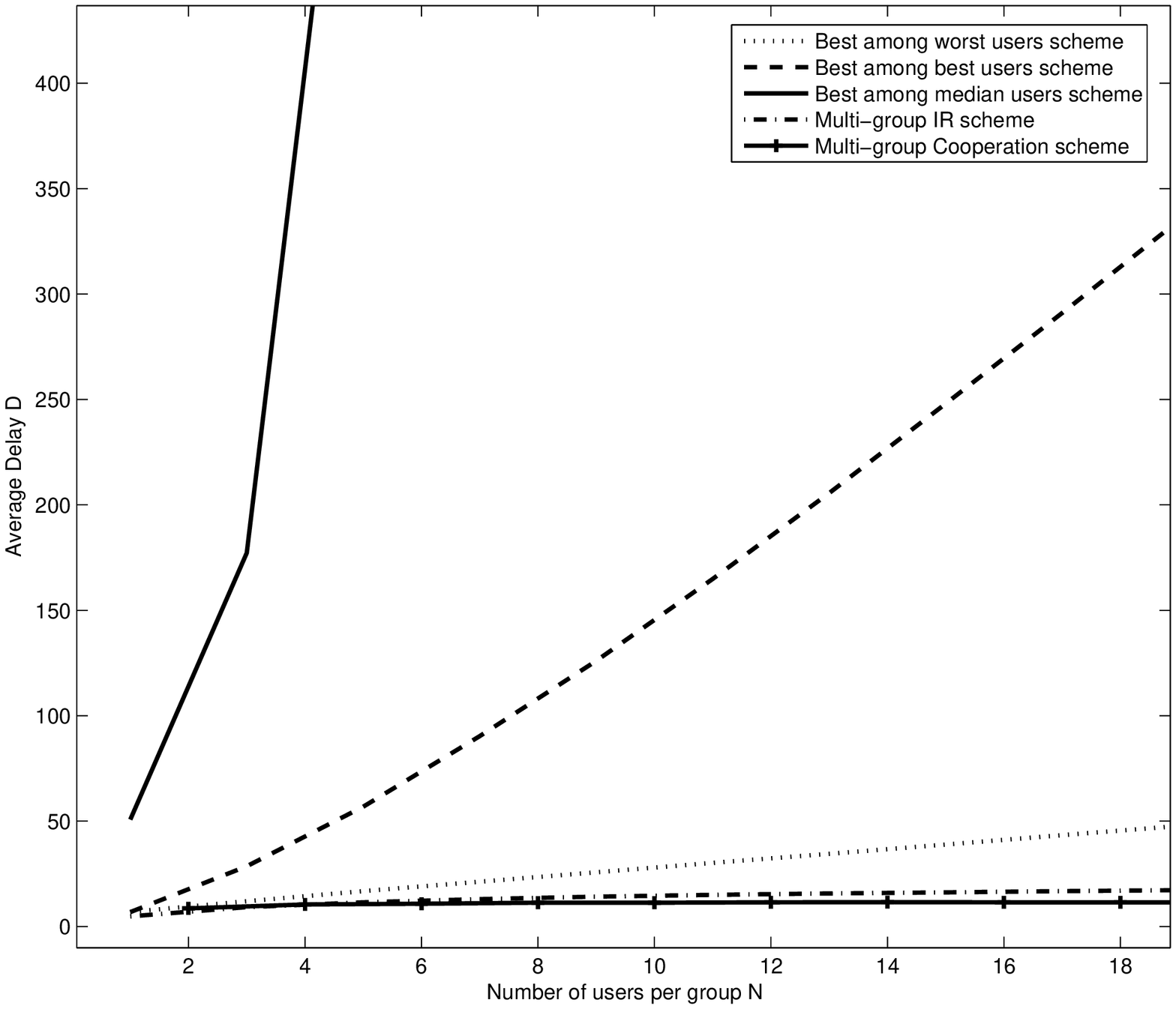}
\caption{Comparison of the delay of the proposed schemes for $G=5$
groups \label{d5}}
\end{figure}

Here we present simulation results that validate our theoretical
claims. These results were obtained through Monte-Carlo
simulations and were averaged over at least 5000 iterations. The
power constraint $P$ is taken to be unity. The throughput of the
static schedulers, proposed in Section~\ref{singlegen}, is shown
in Fig.~\ref{tpos} for different positions of the intended user in
the ordered list of SNRs of all users. It is evident from the
figure that, as predicted by the analysis, the throughput of the
median user scheme is better than that of the best user scheme,
which in turn outperforms the worst user
scheme. In Fig.~\ref{compt}, we present a throughput-comparison
for all the schemes proposed in Section~\ref{singleg} for
increasing values of $N$. The corresponding delay-comparison is
presented in Fig.~\ref{compd}. The throughput-comparison for the
different scheduling schemes in the multi-group scenario is
presented in Fig.~\ref{t5} with $G=5$ groups (the corresponding
delay-comparison is presented in Fig.~\ref{d5}). Although the best
among worst users scheduler performs better than the best among
best users scheme, in terms of throughput, for the range of $N$
values shown in the plot, it should be noted that the latter
eventually outperforms the former for large values of $N$ ($N >
600$). Except for this case, in all other considered scenarios, we
can see that the simulation results follow the same trends
predicted by our asymptotic analysis. Finally, we observe that the
utility of our asymptotic analysis is manifested in its accurate
predictions even with the relatively small number of users used in
our simulations (i.e., in the order of $N=10$).

\section{Conclusions} \label{conc}

In this paper, we have used a cross layer design approach to shed
more light on the throughput-delay tradeoff in the cellular multicast
channel. Towards this end, we proposed three classes of scheduling
algorithms with progressively increasing complexity, and analyzed
the throughput-delay tradeoff achieved by each class. We first
considered the class of low-complexity static scheduling schemes
with memoryless decoding. We showed that a special case of this
scheduling strategy, i.e., the median user scheduler, achieves the
optimal scaling law of the throughput at the expense of an
exponentially increasing delay with the number of users. We then
proposed an incremental redundancy multicast scheme that achieves
a superior throughput-delay tradeoff, at the expense of increased
encoding/decoding complexity. We further proposed a cooperation
scheme that achieves the optimal scaling laws of both throughput
and delay at the expense of a high RF and computational
complexity. We then generalized our schemes to the multi-group
scenario and characterized their ability to exploit the
multi-group diversity offered by the wireless channel. Finally, we
presented simulation results that establish the accuracy of the
predictions of our asymptotic analysis in systems with low to
moderate number of users.

\appendix

\section{Proof of Theorem~\ref{gentd}} \label{app1}

The channel gain $|h_{\pi (N-\frac{N}{\alpha}+1)}|^2$ has the
distribution function $F(x)$ given by
\[ F(x) = \sum_{k=(N-\frac{N}{\alpha}+1)}^N {N \choose k} 
  (1-e^{-x})^k e^{-(N-k)x} , \quad x \ge 0. \]
Hence the average throughput of the proposed scheme is given by
\[ R_{tot} = \frac{N}{\alpha} \int_0^{\infty} \log(1+xP) {\mathrm d}F(x) .\]
Integrating by parts and simplifying, we obtain the average throughput
as stated in equation (\ref{genthru}) of the theorem.

We now calculate the average delay of the proposed scheme.
We consider each coherence interval of length $T_c$ as a time slot.
We f\mbox{}irst calculate the probability distribution of the service time
$X$ required for transmitting a packet (of size $S$) when the base
station always services the same queue. The service time $X$ is def\mbox{}ined as
\beq \label{serdef}
X = k T_c, \qquad k \in \{1,2,\ldots\},
\enq
where $k$ is such that
\beq \label{serdef1}
T_c \left( \sum_{i=1}^{k-1} R_{\alpha}^i \right) < S \le  T_c \left(
\sum_{i=1}^{k} R_{\alpha}^i \right).
\enq
Here $R_{\alpha}^i$ represents the service rate in the $i^{th}$ time slot
as given in (\ref{servrate}). The probability distribution of $X$ is given by
\[ \Prob(X=k T_c) = \Prob\left( \sum_{i=1}^{k-1} R_{\alpha}^i < \frac{S}{T_c}
\le \sum_{i=1}^{k} R_{\alpha}^i \right).\]
We let $C=(S/T_c)$ in the sequel. Using the exponential server assumption in
(\ref{exp}) for the service rates $\{R_{\alpha}^i\}$, we have (for $k \ge 1$)
\[ \Prob(X=k T_c)= \int_{0}^C \left( f_{\left(R_{\alpha}^1+ R_{\alpha}^2 +
\ldots + R_{\alpha}^{(k-1)}\right)}(y) dy \right) \Prob(R_{\alpha}^k > C-y)  \]
\[ = \int_{0}^C \frac{e^{-\mu y} \mu^{k-1} y^{k-2}}{(k-2)!} e^{-\mu (C-y)} dy .\]
\beq \label{genserv}
\Rightarrow \Prob(X=k T_c) = \frac{e^{-\mu C}(\mu C)^{k-1}}{(k-1)!},\qquad
 k \in \{ 1,2,\ldots\}.
\enq
Now the average service time $\bar{X}$ is given by
\[ \bar{X} = \sum_{k=1}^{\infty} k T_c \Prob(X=kT_c) =  \sum_{k=1}^{\infty}
k T_c \left(\frac{e^{-\mu C}(\mu C)^{k-1}}{(k-1)!}\right) .\]
\[ \Rightarrow \bar{X} = (1 + \mu C) T_c = T_c + \mu S. \]
Since $G=1$ for the single group scenario, the assumption in (\ref{Tc}) reduces to
\[ T_c = \Theta \left( \frac{1}{\log \log N} \right). \]
From the results in \cite{Hass2}, we know that
\[ E[R_{\alpha}] \le E[R_N] = \Theta( \log \log N). \]
Hence \[ \mu = \frac{1}{E[R_{\alpha}]} = \Omega \left( \frac{1}
{\log \log N} \right), \quad \forall 1 \le \alpha \le N. \]
Thus for all possible values of the parameter $\alpha$, we have
\[ \bar{X} = T_c + \mu S = \Theta (\mu S) =\Theta \left( \frac{1}{E[R_{\alpha}]} \right). \]
Hence it is clear that the assumption on $T_c$ in (\ref{Tc}) ensures
that the average service time $\bar{X}$ is not dominated by the
scaling behaviour of $T_c$.

We now focus on one set of $\alpha$ coupled queues. Any packet
that arrives into this set enters all the $\alpha$ queues within
the set and moreover, the base station services only one of the
${N \choose N/\alpha}$ available queues in any time slot. Note
that $\bar{X}$ was calculated assuming that the base station
always services the same queue. We are interested in determining
the delay involved in successfully transmitting a particular
packet from all the $\alpha$ coupled queues in the set. The actual
delay, as defined in Section~\ref{model}, is the time between the
start of transmission of a packet and the instant when the packet
reaches all the $N$ users in the system. In our analysis, we
assume that the packet of interest is at the head of all the
$\alpha$ queues in the set during the start of transmission. This
assumption thus yields a lower bound on the actual delay.

We characterize the delay based on the observation that our
queuing problem is equivalent to the well-known ``coupon
collector" problem. This observation was made earlier in
\cite{Hass1} where the authors characterized the delay of the
throughput-optimal broadcast scheme. They assumed that the server
(base station) offers a constant service rate which is independent
of the instantaneous channel gains. In our analysis, however, we
have incorporated the effects of rate adaptation. Let $X_1,X_2,
\cdots,X_{\alpha}$ denote the service times (assuming continuous
service), with distribution as given in (\ref{genserv}), required
for transmitting a packet from each of the $\alpha$ queues in the
set. Then the delay of the proposed scheduling scheme is directly
proportional to the minimum number of trials required to ensure
that the f\mbox{}irst queue is served at least $(X_1/T_c)$ times
by the base station, the second queue is served at least
$(X_2/T_c)$ times and so on ...

We lower bound the average delay by calculating the minimum number
of trials $N_t$ required to ensure that all the $\alpha$ queues are served
at least $(X_{min}/T_c)$ times by the base station, where $X_{min} =
\min \{ X_1, X_2, \cdots, X_{\alpha} \}$. We determine the average number of such
required trials $E[N_t|X_{min}]$ using the results derived in \cite{Hass1}.
Since the base station services only one of the ${N \choose N/\alpha}$ queues
in any time slot and the users are symmetric, there is an equal probability that
the base station services any one of the queues. Since we need to consider only
one set of $\alpha$ coupled queues for determining the delay, we consider all the
other queues in the system jointly as one ``dummy" queue called the $(\alpha+1)^{th}$
queue. Now the probabilities $\{p_j\}$ of the server choosing the $j^{th}$ queue is given by
\[ p_1 = \cdots = p_{\alpha} = \frac{1}{{N \choose N/\alpha}} \quad
\textrm{and} \quad p_{\alpha+1} = P_e = 1 - \frac{\alpha}{{N \choose N/\alpha}} . \]
These probabilities $\{p_j\}$ remain constant through all time slots and are
not functions of the instantaneous service rates $\{R_{\alpha}^i\}$ provided by
the base station. The Moment Generating Function (MGF) of the number of trials
required is given by \cite{Hass1}
\[ F_{N_t|X_{min}}(z) = \sum_{i=0}^{\infty} z^i \Prob(N_t > i) = \sum_{i=0}^{\infty} z^i b_i , \]
where $b_i$ is the probability of failure of sending a packet to all the users
in $i$ channel uses. The value of $b_i$ is equal to the polynomial
\[ \left( \frac{x_1}{{N \choose N/\alpha}} + \cdots + \frac{x_{\alpha}}{{N \choose N/\alpha}}
+  P_e x_{\alpha+1} \right)^i \]
evaluated at $x_1 = \cdots = x_{\alpha+1} =1$ after removing the terms that have all
exponents of $x_1, \cdots, x_{\alpha}$ greater than or equal to $(X_{min}/T_c)$
(denoted by the operator $\{ . \}$). Thus the MGF of the number of trials required is given by
\[ F_{N_t|X_{min}}(z) = \sum_{i=0}^{\infty} \frac{z^i}{\left[ {N \choose N/\alpha} \right]^i }
  \left\{ \left( x_1 + \cdots + x_{\alpha} + P_e {N \choose N/\alpha} x_{\alpha+1}
 \right)^i \right\} \]
evaluated at $x_1 = \cdots = x_{\alpha+1} =1$.
But we know that \[ \frac{i! z^i}{\left[ {N \choose N/\alpha} \right]^i } = \frac{{N
\choose N/\alpha}}{z} \int_0^{\infty} e^{-\frac{{N \choose N/\alpha}t}{z}} t^i dt. \]
Using this identity and simplifying, we get
\[ F_{N_t|X_{min}}(z) = \frac{{N \choose N/\alpha}}{z}  \int_0^{\infty} e^{-\frac{{N \choose
 N/\alpha}t}{z}} \left( e^{{N \choose N/\alpha}t} \left[ 1 - \left( 1 - S_{\left(
 \frac{X_{min}}{T_c} \right)}(t)e^{-t} \right)^{\alpha} \right] \right) dt, \]
where \[ S_m(t) = \sum_{i=0}^{m-1} \frac{t^i}{i!}. \]
Hence the average number of trials required $E[N_t|X_{min}]$ is given by \cite{Hass1}
\[ E[N_t|X_{min}] = F_{N_t|X_{min}}(1) = {N \choose N/\alpha} \int_0^{\infty} \left[1 - \left(
 1 - S_{\left( \frac{X_{min}}{T_c} \right)}(t)e^{-t} \right)^{\alpha} \right] dt  =
 {N \choose N/\alpha} E\left[ \max_{1\le i \le \alpha} Y_i \right], \]
where the $Y_i$'s are i.i.d random variables that follow a Chi-square distribution
with $(2X_{min}/T_c)$ degrees of freedom. From the results in \cite{Hass1}, it can be
seen that for such a sequence of random variables $\{Y_i\}$,
\beq \label{myeqn}
E\left[ \max_{1\le i \le \alpha} Y_i \right] = \max \left\{ \Theta(\log \alpha),
      \Theta \left( \frac{X_{min}}{T_c} \right) \right\}.
\enq
Using this result, the average number of trials required is given by
\[ E[N_t|X_{min}] = \max \left\{ \Theta \left( {N \choose N/\alpha} \log \alpha \right),
   \Theta \left( {N \choose N/\alpha} \frac{X_{min}}{T_c} \right) \right\}. \]
Thus the average delay of the general static scheduling scheme can be lower
bounded by
\[ D \ge E_{X_{min}} \left[ E[N_t|X_{min}] T_c \right] \quad = E_{X_{min}} \left[ \max
 \left\{ \Theta \left( {N \choose N/\alpha} T_c \log \alpha \right), \Theta
 \left( {N \choose N/\alpha} X_{min} \right) \right\} \right]. \]
Since $E\left[ \max \{ Z_1,Z_2 \} \right] \ge \max \left\{ E[Z_1],
E[Z_2] \right\}$, we have
\[ D = \max \left\{ \Omega \left( {N \choose N/\alpha} T_c \log \alpha \right),
    \Omega \left( {N \choose N/\alpha} E[X_{min}] \right) \right\} .\]
\beq \label{Dgen}
\Rightarrow D = \max \left\{ \Omega \left( {N \choose N/\alpha} \frac{
 \log \alpha}{\log \log N} \right), \Omega \left( {N \choose N/\alpha}
  E[X_{min}] \right) \right\}.
\enq
Moreover, when  $E[X_{min}] = \Theta \left( \bar{X} \right)$, it can be
easily seen that the expression on the right in (\ref{Dgen}) gives the
exact scaling of the average delay $D$, instead of just being a lower
bound for it.

\section{Proof of Lemma~\ref{corw}} \label{app2}

The average throughput of the worst user scheme is given by
\beq \label{mint}
R_{tot} = N \int_{0}^{\infty} \log (1 + xP) N e^{-Nx} {\mathrm d} x
= - N e^{\left( \frac{N}{P} \right)} Ei\left(-\frac{N}{P}\right).
\enq
For large values of $x$, we have
\[ Ei(-x) = \int_{-\infty}^{-x} \frac{e^t}{t} {\mathrm d} t
  = - \frac{e^{-x}}{x}  \left(1 + \epsilon \right), \]
where $\epsilon \to 0$ as $x \to \infty$.\
Using this in (\ref{mint}), we get
\[ R_{tot} = P (1 + \epsilon) = \Theta (1). \]
Letting $\alpha=1$ in (\ref{gendel}) of Theorem~\ref{gentd}, we get the
average delay as\footnote{Note that $\Omega \left( {N \choose N/\alpha} \frac{
\log \alpha}{\log \log N} \right) \ne 0$ when $\alpha=1$, since
for any constant $k$, $k + \log \alpha = \Theta (\log \alpha)$  $\forall
1 \le \alpha \le N$.}
\beq \label{dmax}
D = \max \left\{ \Omega \left( \frac{1}{\log \log N} \right),
   \Omega \left( E[X_{min}] \right) \right\}.
\enq
Since the base station maintains only a single queue for the worst user scheme,
we have
\[ E[X_{min}] = \bar{X} = \Theta \left( \frac{1}{E[R_1]} \right). \]
The average service rate $E[R_1]$ is given by
\[ E[R_1] = \int_{0}^{\infty} \log (1 + xP) N e^{-Nx} {\mathrm d} x = -e^{\left(
  \frac{N}{P} \right)} Ei\left(-\frac{N}{P}\right) = \Theta
  \left( \frac{1}{N} \right). \]
Since $E[X_{min}] = \bar{X}$, the expression on the
right in (\ref{dmax}) gives the exact scaling of $D$. Thus the average
delay of the worst user scheme scales as $D=\Theta(N)$.

\section{Proof of Lemma~\ref{corb}} \label{app3}

By letting $\alpha=N$ in (\ref{genthru}) of Theorem~\ref{gentd}, the average
throughput of the best user scheme is found to be
\beq \label{maxt}
R_{tot} = \sum_{i=1}^N {N \choose i} (-1)^i e^{\left(\frac{i}{P}\right)}
 Ei\left(\frac{-i}{P}\right).
\enq
It has been shown in \cite{Hass2} that the throughput in (\ref{maxt}) scales as
\beq \label{maxscal}
R_{tot} = \Theta(\log \log N)
\enq
with the number of users $N$. Hence the average service rate is given by
\[ E[R_N] = R_{tot} = \Theta \left( \log \log N \right). \]
Letting $\alpha=N$ in (\ref{gendel}) of Theorem~\ref{gentd}, we get the
average delay as
\beq \label{dmin}
D = \max \left\{ \Omega \left( \frac{N\log N}{\log \log N} \right),
   \Omega \left( N E[X_{min}] \right) \right\}.
\enq
Now \[ E[X_{min}] = E\left[\min_{i=1}^{N} X_i \right] \le \bar{X} =
  \Theta \left( \frac{1}{E[R_{N}]} \right). \]
Hence \[ E[X_{min}] = O \left( \frac{1}{\log \log N} \right). \]
Thus, from (\ref{dmin}), the average delay of the best user scheme
scales as
\[ D = \Omega \left( \frac{N \log N}{\log \log N} \right). \]

\section{Proof of Lemma~\ref{corm}} \label{app4}

The average throughput of the median user scheme can
be derived by letting $\alpha=2$ in (\ref{genthru}) of
Theorem~\ref{gentd}. From the results on central order statistics
in \cite{book} (Theorem~8.5.1), we know that the sample
median of $N$ i.i.d. exponential random variables converges in
distribution to a normal random variable with mean $\theta$ and
variance $(1/N)$, where $\theta = \log 2$ is the median of
the underlying exponential distribution. Hence
\[ \left( |h_{\pi (\frac{N}{2}+1)}|^2 - \theta \right) \sqrt{N}
    \to W \quad \textrm{in distribution}, \]
where $W$ is a standard normal random variable. Using Chebyshev's
inequality, we get $\forall \epsilon >0$
\[ \Prob \left( \left| |h_{\pi (\frac{N}{2}+1)}|^2 - \theta \right|
  > \epsilon \right) = \Prob \left( \sqrt{N} \left| |h_{\pi (\frac{N}{2}
  +1)}|^2 - \theta \right| > \epsilon \sqrt{N} \right) < \frac{ E[W^2]
 + \delta }{N \epsilon^2} \to 0 \quad \textrm{as $N \to \infty$}. \]
\[ \Rightarrow  |h_{\pi (\frac{N}{2}+1)}|^2 \to \theta \quad
 \textrm{in probability}. \]
Since the $\log(.)$ function is continuous, we have
\beq \label{pcon}
\log \left( 1 + |h_{\pi (\frac{N}{2}+1)}|^2 P \right) \to
\log (1 + \theta P) \quad  \textrm{in probability}.
\enq
We now derive a lower bound on the average throughput. We recall
the following property of positive random variables.
Let $(X_n)$ be a set of positive random variables converging to
a constant $A$ in probability. Hence $\forall \epsilon >0$,
\[ \Prob \left( |X_n - A| \ge \epsilon \right) < \delta, \]
for some small $\delta >0$. Now
\[ E[X_n] = \int_{0}^{\infty} t f_{X_n}(t) {\mathrm d} t \quad \ge \int_{A-
  \epsilon}^{A+\epsilon} t f_{X_n}(t) {\mathrm d} t \quad \ge (A-\epsilon)
  (1 - \delta). \]
Taking the limit as $n \to \infty$, we get
\beq \label{posrv}
\lim_{n \to \infty} E[X_n] \ge A.
\enq
Using this property in (\ref{pcon}), we get
\[ \lim_{N \to \infty} E\left[ \log \left( 1 + |h_{\pi (\frac{N}{2}
  +1)}|^2 P \right) \right] \ge \log (1 + \theta P) = \Theta(1) .\]
\beq \label{lobound}
\Rightarrow R_{tot} = \left(\frac{N}{2}\right) E\left[ \log \left(
   1 + |h_{\pi (\frac{N}{2}+1)}|^2 P \right) \right] = \Omega(N).
\enq
An upper bound on the average throughput of {\bf any} scheduling scheme
is given by
\[ R_{tot} \le E \left[\sum_{i=1}^N \log \left( 1 + |h_i|^2 P \right)
   \right] = N E \left[ \log \left( 1 + |h_1|^2 P \right) \right]. \]
Hence
\beq \label{upbound}
R_{tot} = O(N).
\enq
Combining this with the lower bound in (\ref{lobound}), we get
\[ R_{tot} = \Theta (N). \]
Thus it is clear that the throughput of the proposed median user scheme
is scaling law optimal. Letting $\alpha=2$ in (\ref{gendel}) of
Theorem~\ref{gentd}, we get the average delay as
\beq \label{dmed}
D = \max \left\{ \Omega \left( {N \choose N/2}\frac{1}{\log \log N} \right),
   \Omega \left({N \choose N/2} E[X_{min}] \right) \right\}.
\enq
Now, since $X_{min} = \min \{ X_1,X_2 \}$, we have
\[ E[X_{min}] =  \Theta \left( \bar{X} \right) = \Theta \left(
   \frac{1}{E[R_2]} \right). \]
The average service rate $E[R_2]$ is given by
\beq \label{avgmed}
E[R_2] = E\left[ \log \left( 1 + |h_{\pi (\frac{N}{2}+1)}|^2 P \right)
\right] = \Theta (1).
\enq
Since $E[X_{min}] =  \Theta \left( \bar{X} \right)$, the expression on the
right in (\ref{dmed}) gives the exact scaling of $D$.
Thus the average delay of the median user scheme scales as
\[ D = \Theta \left( {N \choose N/2} \right) .\]
Using Stirling's formula, we obtain the scaling of the average delay
as given in (\ref{meddel}).

\section{Proof of Theorem~\ref{thinc}} \label{app5}

Let $A_i$ denote the event that a packet is successfully decoded by all
the $N$ users in the system in $i$ transmission attempts.
Following the notation in \cite{Caire}, we def\mbox{}ine
\[ q(m) = \Prob(\overline{A_1},\ldots,\overline{A_{m-1}},A_m) = p(m-1) - p(m), \]
where \[ p(m) = \Prob(\overline{A_1},\ldots,\overline{A_{m-1}},\overline{A_m})
 = 1-\sum_{l=1}^m q(l), \]
with $p(0)=1$. The rate $\bar{R}$ is def\mbox{}ined as $\bar{R}=(b/L)$.
We def\mbox{}ine the random variable $\tau$ to be the number of transmission
attempts made between the instant when the codeword is generated and the instant
when its transmission is stopped (Transmission is stopped either when the
packet is successfully decoded by all the $N$ users or the number of
transmission attempts exceeds the rate constraint $M$). The probability
distribution of $\tau$ is given by
\[ f_{\tau}(m) = \left\{ \begin{array}{l} 0, \quad m=0 \\ q(m), \quad
  1\le m \le M-1 \\ q(M) +p(M), \quad m=M \end{array} \right. \]
We def\mbox{}ine the random reward ${\mathcal R}$ as follows:
${\mathcal R} = N \bar{R}$ if transmission stops because of successful
decoding and ${\mathcal R} =0$ if transmission stops because of the
rate constraint violation.
Hence
\[ E[{\mathcal R}] = N \bar{R} \sum_{m=1}^{M} q(m) = N \bar{R} [1-p(M)]. \]
The mean inter-renewal time is given by
\[ E[\tau] = \sum_{m=1}^M m f_{\tau}(m) = \sum_{m=1}^M m q(m) + Mp(M)
  = \sum_{m=1}^M m[p(m-1) - p(m)] + Mp(M) = \sum_{m=0}^{M-1} p(m) . \]
Applying the renewal-reward theorem, we obtain the average throughput
of the proposed scheme as
\[ R_{tot} = \frac{E[{\mathcal R}]}{E[\tau]}, \quad \textrm{with
  probability $1$}. \]
Hence \[ R_{tot} = \frac{N \bar{R} \left[ 1-p (M) \right]}{1 +
  \sum_{m=1}^{M-1} p(m)} .\]
The average delay $D$ of the scheme is given by the mean inter-renewal
time. Hence
\[ D = E[\tau] = 1 + \sum_{m=1}^{M-1} p(m). \]
The unconstrained throughput and delay of this scheme are obtained by
letting $M \to \infty$ and are given by
\beq
R_{tot} = \frac{N \bar{R}}{\sum_{m=0}^{\infty} p(m)},
\enq
and
\beq
D = \sum_{m=0}^{\infty} p(m).
\enq
From the earlier def\mbox{}initions, we have
\[ p(m) = \Prob(\overline{A_1},\ldots,\overline{A_{m-1}},\overline{A_m}) =
\Prob(\overline{A_m}) = \Prob\left(\min_{i=1}^N \sum_{k=1}^m I(X;Y_{ik}) \le \bar{R}
\right) .\]
\beq \label{pmeq}
\Rightarrow p(m) =  1 - \Prob\left(\min_{i=1}^N \sum_{k=1}^m I(X;Y_{ik}) > \bar{R}
 \right) = 1 - \left[ 1 - \Prob\left( \sum_{k=1}^m I(X;Y_{1k}) \le \bar{R}
\right) \right]^N.
\enq
Now for a Gaussian input distribution, we have
\[ \sum_{k=1}^m I(X;Y_{1k}) = \sum_{k=1}^m \log (1 + |h_{k}|^2 ). \]
We know that
\[ \log \left( 1 + \sum_{k=1}^m |h_k|^2 \right) \le \sum_{k=1}^m
\log (1 + |h_{k}|^2 ) \le \sum_{k=1}^m |h_k|^2. \]
Hence
\[ \Prob\left(\sum_{k=1}^m |h_k|^2 \le (e^{\bar{R}}-1) \right) \ge 
\Prob\left( \sum_{k=1}^m \log (1 + |h_{k}|^2 ) \le \bar{R} \right) \ge
\Prob \left( \sum_{k=1}^m |h_k|^2 \le \bar{R} \right) . \]
Since both $\bar{R}$ and $(e^{\bar{R}}-1)$ are constants, substituting both
the lower and upper bounds in (\ref{pmeq}) will yield the same scaling with
$N$. So we consider only the lower bound on $p(m)$. Let
\[ s(m) = 1 - \left[ 1 - \Prob\left(\sum_{k=1}^m |h_k|^2 \le \bar{R} \right)
  \right] ^N. \]
Hence
\[ \sum_{m=0}^{\infty} p(m) = \Theta \left( \sum_{m=0}^{\infty} s(m)
\right) \quad \textrm{w.r.t $N$}. \]
The random variable $\sum_{k=1}^m  |h_k|^2$ has a
$2m$-dimensional Chi-square distribution with the density and 
distribution functions given by
\[ f(x) = \frac{e^{-x} x^{m-1}}{(m-1)!} , \quad x \ge 0 \] and
\[F(x) = 1 - e^{-x} \left( \sum_{l=0}^{m-1} \frac{x^l}{l!} \right), \quad
 x \ge 0. \]
Hence
\[ s(m) = 1 - \left[ e^{-\bar{R}} \left( \sum_{l=0}^{m-1} \frac{\bar{R}^l}
  {l!} \right) \right]^N. \]
From Taylor's theorem, we know that (for some $0 < \theta < 1$)
\[ e^{\bar{R}} = \sum_{l=0}^{m-1} \frac{\bar{R}^l}{l!} +  \frac{e^{\theta
 \bar{R}} {\bar{R}}^m}{m!} \quad \Rightarrow \sum_{l=0}^{m-1} \frac{
 \bar{R}^l}{l!} = e^{\bar{R}} - \frac{e^{\theta \bar{R}} {\bar{R}}^m}{m!}. \]
\[ \Rightarrow s(m) = 1 - \left( 1 - \frac{e^{-(1-\theta )\bar{R}}
   \bar{R}^m}{m!} \right)^N. \]
To f\mbox{}ind the scaling of $\sum_{m=0}^{\infty} s(m)$ w.r.t $N$, we f\mbox{}irst derive
a lower bound by f\mbox{}inding the value of $m$ until which $s(m) \to 1$ as $N \to
\infty$. Now \[ s(m) \to 1 \quad \Rightarrow \left( 1 - \frac{e^{-(1-\theta )
\bar{R}} \bar{R}^m}{m!} \right)^N \to 0 .\]
\[ \Rightarrow \frac{e^{-(1-\theta )\bar{R}} \bar{R}^m}{m!} > \Theta \left( \frac{1}{N}
\right). \]
Using Stirling's approximation, we have
\[ \frac{e^{-(1-\theta )\bar{R}} \bar{R}^m}{\sqrt{2 \pi m}  e^{-m} m^m} > \frac{k}{N},
 \quad \forall \quad \textrm{constant $k$}. \]
Taking log on both sides, we get
\[ (1-\theta )\bar{R} - m + m \log \left( \frac{m}{\bar{R}} \right) + \frac{1}{2}
\log (2 \pi m ) < \log N - \log k, \quad \forall k.\]
For large $N$, this equation can be reduced to
\beq \label{cond}
m \log m < \log N .
\enq
This equation is satisf\mbox{}ied by all values of $m$ such that
\[ m < \Theta \left( \frac{\log N}{\log \log N} \right). \]
Since $s(m) \to 1$ as $N \to \infty$ for all values of $m$ that
satisfy the above equation, the sum of $s(m)$'s can be lower
bounded as
\beq  \label{inclow}
\sum_{m=0}^{\infty} s(m) \ge \Theta \left( \frac{\log N}
{\log \log N} \right).
\enq
Similarly an upper bound on $\sum_{m=0}^{\infty} s(m)$ can be derived by
f\mbox{}inding the value of $m$ from which $s(m) \to 0$ as $N \to \infty$.
Following the same procedure as before, we f\mbox{}ind that $s(m) \to 0$ when
\[ m > \Theta \left( \frac{\log N}{\log \log N} \right). \]
This yields the following upper bound
\[ \sum_{m=0}^{\infty} s(m) \le \Theta \left( \frac{\log N}
  {\log \log N} \right). \]
Combining this with the lower bound in (\ref{inclow}), we get
\[ \sum_{m=0}^{\infty} s(m) = \Theta \left( \frac{\log N}
  {\log \log N} \right). \]
Thus the average delay is given by
\[ D= \sum_{m=0}^{\infty} p(m) = \Theta \left( \sum_{m=0}^{\infty}
 s(m) \right) = \Theta \left( \frac{\log N}{\log \log N} \right). \]
The average throughput of the incremental redundancy scheme is then given by
\[ R_{tot} = \frac{N \bar{R}}{\sum_{m=0}^{\infty} p(m)} = \frac{N \bar{R}}{D}
 = \Theta \left( \frac{N \log \log N}{\log N} \right). \]

\section{Proof of Theorem~\ref{coop}} \label{app6}

The f\mbox{}irst stage of the cooperation scheme is the median user scheme.
Hence it is clear from (\ref{medscal}) that
\[ E[R_{s1}] = E\left[ \log \left( 1 + |h_{\pi (\frac{N}{2}+1)}|^2 P\right) \right]
 = \Theta(1).  \]
As noted earlier, the cooperative transmission by the users in the second stage
is equivalent to the transmission of packets from a transmitter equipped with
$(N/2)$ transmit antennas to the worst user in a group of $(N/2)$ users. Hence
the average transmission rate during the cooperative stage is given by
\[ E[R_{s2}] = E\left[ \min_{i=1,...,(N/2)} \log \left( 1 + \frac{|h_{1i}|^2
 + \cdots + |h_{(N/2)i}|^2}{(N/2)} P \right) \right], \]
where the $|h_{ki}|^2$'s are i.i.d and exponentially distributed and represent
the inter-user fading coeff\mbox{}icients.
\begin{equation} \label{Req}
\Rightarrow E[R_{s2}] = E\left[ \log \left( 1 + \min_{i=1,...,M}
\frac{|\chi^i_{2M} |^2}{M} P \right) \right],
\end{equation}
where $M=(N/2)$ and $|\chi^i_{2M} |^2$'s are Chi-square random variables with
$2M$ degrees of freedom whose distribution function is given by
\[ F(x) = 1 - e^{-x} \left( \sum_{j=0}^{M-1} \frac{x^j}{j!} \right), \quad x \ge 0. \]
Using the results on extreme order statistics in \cite{book} (Theorems~8.3.2-8.3.6),
it can be shown that the random variable
\[ \frac{ \min_{i=1}^M |\chi^i_{2M} |^2 }{b_M} \quad \to \quad W \quad
 \textrm{in distribution as $M \to \infty$}, \]
where $W$ is a Weibull type random variable and $b_M$ satisf\mbox{}ies
$F(b_M) = \frac{1}{M}$. Now
\[ F(b_M) = \frac{1}{M} \quad \Rightarrow 1 - e^{-b_M} \left(
\sum_{j=0}^{M-1} \frac{b_M^j}{j!} \right) = \frac{1}{M}. \]
Using Taylor's theorem, we get for some $0< \beta_M <1$
\[ 1 - e^{-b_M} \left( e^{b_M} - \frac{e^{\beta_M b_M} b_M^M}{M!} \right) = \frac{1}{M}
  \quad \Rightarrow \frac{ e^{-(1-\beta_M) b_M} b_M^M}{M!} = \frac{1}{M} .\]
Using Stirling's approximation, we have \[ \frac{e^{-(1-\beta_M) b_M} b_M^M}{\sqrt{2\pi M}
M^M e^{-M}} = \frac{1}{M}. \]
Taking $\log(.)$ on both sides, we get
\[ (1-\beta_M) b_M - M \log b_M = M - \left( M-\frac{1}{2} \right) \log M + C . \]
Since $\beta_M \to 0$ as $M \to \infty$, we get $b_M = \Theta (M)$.
Thus \[ \frac{ \min_{i=1}^M |\chi^i_{2M} |^2 }{M} \to kW \quad
\textrm{in distribution, for some constant $k>0$}. \]
Since the $\log(.)$ function is continuous, we have
\[ \log \left( 1 + \frac{ \min_{i=1}^M |\chi^i_{2M} |^2 }{M} P \right) \to
   \log( 1 + kWP) \quad \textrm{in distribution, as $M \to \infty$}. \]
Now, we know
\[ \log \left( 1 + \frac{ \min_{i=1}^M |\chi^i_{2M} |^2 }{M} P \right) \le
  \frac{ \left(\min_{i=1}^M |\chi^i_{2M} |^2 \right) P}{M} \le \frac{
  |\chi^1_{2M} |^2 P}{M}. \]
Since \[ E\left[ \left( \frac{|\chi^1_{2M} |^2 P}{M} \right)^2 \right] =
  \frac{E[(|\chi^1_{2M} |^2)^2] P^2}{M^2} = \left( \frac{M^2 + M}{M^2} \right)
  P^2 = \left( 1 + \frac{1}{M}\right) P^2 \le 2P^2 < \infty \quad \forall M, \]
\[ \Rightarrow  \left\{ \frac{|\chi^1_{2M} |^2 P}{M} ; M \ge 1 \right\} \quad
\textrm{is uniformly integrable.} \]
\[ \Rightarrow  \left\{ \log\left( 1 + \frac{ \min_{i=1}^M |\chi^i_{2M} |^2 }{M}
 P \right) ; M \ge 1 \right\}
   \quad \textrm{is uniformly integrable.} \]
It is shown in \cite{Durrett} that if a sequence of random variables $(X_n)$ is
uniformly integrable and $X_n \to X$ in distribution as $n \to \infty$, then
$EX_n \to EX$ as $n \to \infty$.
Thus \[ E \left[ \log \left( 1 + \frac{ \min_{i=1}^M |\chi^i_{2M} |^2 }{M} P\right)
 \right] \to E[ \log (1 + kWP)] = \Theta(1). \]
Hence the average transmission rate of the second stage is given
by $E[R_{s2}] = \Theta(1)$ w.r.t $N$. Since both $E[R_{s1}]$
and $E[R_{s2}]$ do not scale with $N$ and since the minimum is
taken over only two positive quantities, we have \[ E \left[\min
\{ R_{s1} , R_{s2} \} \right] = \Theta \left( E[R_{s1}] \right) =
\Theta(1).
\] Thus the average throughput of the cooperation scheme is given
by
\[ R_{tot} = \left( \frac{N}{2} \right) E \left[\min \{ R_{s1} , R_{s2} \}\
  \right] = \Theta(N). \]
We now determine the average delay of the cooperation scheme. We note that
the base station needs to maintain only a single queue that caters to all
the $N$ users in the system. The information transmitted by the base station
in the f\mbox{}irst half of each time slot reaches all the $N$ users at the end of
that time slot. Hence the average delay is equal to the average service time
required for transmitting a packet of size $S$ from the queue. Following the
steps in Appendix~\ref{app1}, the average delay $D$ for transmitting a packet
in the cooperation scheme is given by
\[ D = T_c + \mu S = \Theta \left( \frac{S}{E[\min \{ R_{s1}, R_{s2} \}]} \right)
 = \Theta(1). \]

\section{Proof of Theorem~\ref{statmg}} \label{app7}

We f\mbox{}irst extend the proof of the general static scheduling scheme
given in Appendix~\ref{app1} to the multi-group scenario and then
consider the three special cases. The average throughput $R_{tot}$
of the general multi-group scheduling scheme is given by
\[ R_{tot} = \left( \frac{N}{\alpha} \right) E[R_{\alpha}], \]
where $R_{\alpha}$ represents the transmission rate to each of the
intended $(N/\alpha)$ users and is given by
\[ R_{\alpha} = \log \left( 1 + |h_{bg}|^2 P \right), \]
where the distribution of $|h_{bg}|^2$ is given by
\beq \label{bgdis}
F_{bg}(x) = \left( \sum_{k=(N - \frac{N}{\alpha}+1)}^N {N \choose k}
 \left(1-e^{-x} \right)^k e^{-(N-k)x} \right)^G, \forall x\ge 0.
\enq
Hence the average throughput is given by
\beq \label{mgtput}
R_{tot} = \frac{N}{\alpha} \int_0^{\infty} \log(1+xP) {\mathrm d}F_{bg}(x).
\enq
Integrating by parts and simplifying, we obtain the average throughput
of the proposed scheme.

For implementing the general multi-group static scheduling scheme, the base
station needs to maintain $G {N \choose N/\alpha}$ queues, one for
each combination of $(N/\alpha)$ users in each of the $G$ groups. These
queues can be divided into $G$ sets, one for each of the groups. Within
each set corresponding to a particular group, the queues can be further divided
into subsets with $\alpha$ coupled queues in each subset such that the
combinations of users served by the $\alpha$ queues within a subset are
mutually exclusive and collectively exhaustive (i.e., every user in the
particular group is served by exactly one of the $\alpha$ queues). We
consider one such subset of $\alpha$ queues corresponding to any one of
the $G$ groups. Any packet that arrives into the subset enters all the
$\alpha$ queues since it needs to be transmitted to all the users within
the group. At any instant of time, the base station services only one
of the $G {N \choose N/\alpha}$ queues.

As before, we f\mbox{}irst calculate the average service
time $\bar{X}$ required for transmitting a packet by assuming that the base
station always services the same queue. Following the steps in
Appendix~\ref{app1}, the average service time $\bar{X}$ is given by
\[ \bar{X} = T_c + \mu S =\Theta \left( \frac{S}{E[R_{\alpha}]} \right). \]

We again use the results in \cite{Hass1} to derive a lower bound on the
actual delay by considering the minimum number of trials $N_t$ required
to ensure that all the $\alpha$ queues are served at least $(X_{min}/T_c)$
times by the base station. As before, we consider only $(\alpha+1)$ queues
with the $(\alpha+1)^{th}$ queue being the ``dummy" queue representing
all the queues in all other subsets in the system. Now the probabilities
$\{p_i\}$ of the server choosing the $i^{th}$ queue are given by
\[ p_1 = \cdots = p_{\alpha} = \frac{1}{G {N \choose N/\alpha}} \quad
\textrm{and} \quad p_{\alpha+1} = 1 - \frac{\alpha}{G {N \choose N/\alpha}} . \]
Proceeding as in Appendix~\ref{app1}, the MGF of the number of trials required is given by
\[ F_{N_t|X_{min}}(z) = \frac{G {N \choose N/\alpha}}{z}  \int_0^{\infty}
 e^{-\frac{G {N \choose N/\alpha}t}{z}} \left( e^{G {N \choose N/\alpha}t}
 \left[ 1 - \left( 1 - S_{\left(\frac{X_{min}}{T_c}\right)}(t)e^{-t}
 \right)^{\alpha} \right] \right) dt. \]
The average number of trials required $E[N_t|X_{min}]$ is given by
\[ E[N_t|X_{min}] = F_{N_t|X_{min}}(1) = G {N \choose N/\alpha} \int_0^{\infty} \left[1 -
 \left(1 - S_{\left(\frac{X_{min}}{T_c}\right)}(t)e^{-t}\right)^{\alpha} \right] dt
  = G {N \choose N/\alpha} E\left[ \max_{1\le i \le \alpha} Y_i \right], \]
where the $Y_i$'s are i.i.d random variables that follow a Chi-square distribution
with $(2X_{min}/T_c)$ degrees of freedom. Using the result in (\ref{myeqn}), we get
\[ E[N_t|X_{min}] =  \max \left\{ \Theta \left( G {N \choose N/\alpha}\log \alpha \right),
\Theta \left( G {N \choose N/\alpha} \frac{X_{min}}{T_c} \right) \right\}. \]
Thus the average delay of the proposed scheme can be lower bounded as
\[ D \ge E_{X_{min}} \left[ E[N_t|X_{min}] T_c \right] = E_{X_{min}} \left[ \max
 \left\{ \Theta \left( G {N \choose N/\alpha} T_c \log \alpha \right), \Theta
 \left( G {N \choose N/\alpha} X_{min} \right) \right\} \right]. \]
\beq \label{mgd}
\Rightarrow D = \max \left\{ \Omega \left( {N \choose N/\alpha} \frac{G \log
 \alpha}{\log \log NG} \right), \Omega \left( G {N \choose N/\alpha} E[X_{min}] \right) \right\}.
\enq
Moreover, when  $E[X_{min}] = \Theta \left( \bar{X} \right)$, the
expression on the right in (\ref{mgd}) gives the exact scaling of
the average delay $D$, instead of just being a lower bound for it.

\subsection{Best among worst users scheme ($\alpha=1$)}
Letting $\alpha=1$ in (\ref{mgtput}) and simplifying,
we get the average throughput to be
\beq \label{mintmg}
R_{tot} = N \left[ \sum_{k=1}^G {G \choose k} (-1)^k e^{\left( \frac{Nk}{P}
 \right)} Ei\left(-\frac{Nk}{P}\right) \right].
\enq
For large values of $x$, we have
\[ Ei(-x) = \int_{-\infty}^{-x} \frac{e^t}{t} {\mathrm d} t = - \frac{e^{-x}}{x}
 \left(1 + \epsilon \right), \]
where $\epsilon \to 0$ as $x \to \infty$. Using this in (\ref{mintmg}), we get
\[ R_{tot} = P \left[ \sum_{k=1}^G {G \choose k} \frac{(-1)^{k+1}}{k} \right]
 (1+ \epsilon) . \]
It can be shown using the results in \cite{intbook} that
\[ R_{tot} = P \left[ \sum_{k=1}^G \frac{1}{k} \right] (1 + \epsilon)
 = \Theta ( \log G). \]
The average service rate is given by
\[ E[R_1] = \sum_{k=1}^G {G \choose k} (-1)^k e^{\left( \frac{Nk}{P} \right)}
 Ei\left(-\frac{Nk}{P}\right) = \Theta \left( \frac{\log G}{N} \right). \]
Letting $\alpha=1$ in (\ref{mgd}) and using the fact that
\[ E[X_{min}] = \bar{X} = \Theta \left( \frac{1}{E[R_1]} \right), \]
we get the average delay as
\[ D = \Theta \left( \frac{NG}{\log G} \right). \]

\subsection{Best among best users scheme ($\alpha=N$)}
Letting $\alpha=N$ in (\ref{mgtput}) and simplifying, we get
the average throughput to be
\beq \label{maxtmg}
R_{tot} = \sum_{k=1}^{NG} {NG \choose k} (-1)^k e^{\left( \frac{k}{P}
 \right)} Ei\left(-\frac{k}{P}\right).
\enq
It has been shown in \cite{Hass2} that the throughput in (\ref{maxtmg}) scales as
\beq \label{maxscalmg}
R_{tot} = \Theta(\log \log NG).
\enq
Hence the average service rate is given by
\[ E[R_N] = R_{tot} = \Theta \left( \log \log NG \right). \]
Letting $\alpha=N$ in (\ref{mgd}) and using the fact that
\[ E[X_{min}] \le \bar{X} = \Theta \left( \frac{1}{E[R_{N}]} \right), \]
we get the average delay as
\[ D = \Omega \left( \frac{NG \log N}{\log \log NG} \right). \]

\subsection{Best among median users scheme ($\alpha=2$)}
The average throughput of the best among median users scheme can be derived
by letting $\alpha=2$ in (\ref{mgtput}). It is given by
\[ R_{tot} = \left(\frac{N}{2}\right) E\left[ \log \left( 1 + \max_{g=1}^G
 |h_{\pi (\frac{N}{2}+1)}^g|^2 P \right) \right] . \]
We now determine bounds on the asymptotic scaling of $R_{tot}$ as $N$ and $G$ grow
to inf\mbox{}inity.
To get a lower bound on the throughput, we use the fact that $\max \{ X_1,
\cdots, X_n \} \ge X_1$ and obtain
\[ R_{tot} \ge \left(\frac{N}{2}\right) E\left[ \log \left( 1 +
 |h_{\pi (\frac{N}{2}+1)}^1|^2 P \right) \right] . \]
The expression on the right is clearly the throughput of the median user
scheduler described in Section~\ref{singlegen}. Hence, from Lemma~\ref{corm},
we get \[ R_{tot} = \Omega (N). \]
We derive an upper bound on the throughput by using the fact that for
continuous unimodal distributions
\[ \left| \textrm{Median} - \textrm{Mean} \right| < \textrm{Standard Deviation}. \]
Hence\footnote{It is easy to show using convergence arguments that the
inequality is valid for the empirical values used in the proof.}
\[ |h_{\pi (\frac{N}{2}+1)}|^2 < \frac{|h_1|^2 + \cdots + |h_N|^2}{N} +
 \sqrt{ \left( \frac{|h_1|^4 + \cdots + |h_N|^4}{N} \right) - \left(
  \frac{|h_1|^2 + \cdots + |h_N|^2}{N} \right)^2 } .\]
\[ \Rightarrow E \left[ \max_{g=1}^G |h_{\pi (\frac{N}{2}+1)}^g|^2 \right] <
  E \left[ \max_{g=1}^G \left\{ \frac{|h_1^g|^2 + \cdots + |h_N^g|^2}{N} +
 \sqrt{  \frac{|h_1^g|^4 + \cdots + |h_N^g|^4}{N} } \right\} \right]. \]
Using Jensen's inequality and the fact that $\max_{i=1}^n \{X_i + Y_i\}
 \le \max_{i=1}^n \{X_i\} + max_{i=1}^n \{Y_i \}$, we get
\[ E \left[ \max_{g=1}^G |h_{\pi (\frac{N}{2}+1)}^g|^2 \right] < E \left[
 \max_{g=1}^G \left\{ \frac{|h_1^g|^2 + \cdots + |h_N^g|^2}{N} \right\}
 \right] + \sqrt{ E \left[ \max_{g=1}^G \left\{ \frac{|h_1^g|^4 + \cdots +
 |h_N^g|^4}{N} \right\} \right] } \]
\[ < \frac{1}{N} \sum_{i=1}^N E \left[ \max_{g=1}^G |h_i^g|^2 \right] +
  \sqrt{ \frac{1}{N} \sum_{i=1}^N E \left[ \max_{g=1}^G |h_i^g|^4 \right]}
 = E \left[ \max_{g=1}^G |h_1^g|^2 \right] + \sqrt{ E \left[ \left(
  \max_{g=1}^G |h_1^g|^2 \right)^2 \right] }. \]
It is known that for a sequence of exponential random variables $\{X_i\}$
with unit mean \cite{book},
\[ E \left[ \max_{i=1}^G X_i \right] = \Theta (\log G) \quad \textrm{and}
\quad E \left[ \left( \max_{i=1}^G X_i \right)^2 \right] = \Theta \left(
(\log G)^2 \right). \]
Thus \[ E \left[ \max_{g=1}^G |h_{\pi (\frac{N}{2}+1)}^g|^2 \right] =
O (\log G). \]
Now applying Jensen's inequality, we get the upper bound on the average throughput as
\[ R_{tot} = \left(\frac{N}{2}\right) E\left[ \log \left( 1 + \max_{g=1}^G
 |h_{\pi (\frac{N}{2}+1)}^g|^2 P \right) \right] \le \left(\frac{N}{2}\right)
 \log \left( 1 + E\left[ \max_{g=1}^G |h_{\pi (\frac{N}{2}+1)}^g|^2 \right] P
 \right) .\]
\[ \Rightarrow R_{tot} = O (N \log \log G). \]
Thus the average throughput of the best among median users scheme can be
bounded as
\beq
\Omega (N) = R_{tot} = O( N \log \log G).
\enq
The average service rate $E[R_2]$ is then bounded by
\[ \Omega (1) = E[R_2] = O( \log \log G). \]
Letting $\alpha=2$ in (\ref{mgd}) and using the fact that
\[ E[X_{min}] = \Theta \left( \bar{X} \right) = \Theta \left(
   \frac{1}{E[R_2]} \right), \]
the average delay can be bounded as
\[ \Omega \left( {N \choose N/2}\frac{G}{\log \log G} \right) =  D
  = O \left( {N \choose N/2} G \right) .\]
Using Stirling's formula, we obtain bounds on the average delay as stated in
(\ref{meddelmg}).

\section{Proof of Theorem~\ref{coopmg}} \label{app8}

The average throughput of the multi-group cooperation scheme is given by
\[ R_{tot} = E \left[ \max_{g=1}^G \left\{ \left( \frac{N}{2} \right)
   \min \{ R_{s1}^g , R_{s2}^g \} \right\} \right]. \]
Since $E[ \max \{X_1,\cdots,X_n\}] \ge E[X_1]$, we have
\[ R_{tot} \ge E \left[ \left( \frac{N}{2} \right) \min \{ R_{s1}^1 ,
R_{s2}^1 \} \right]. \]
The expression on the right is the average throughput of the single group
cooperation scheme described in Section~\ref{singlecoop}. Using the
results of Theorem~\ref{coop}, we have
\[ R_{tot} = \Omega(N). \]
The average throughput can be upper bounded by using the fact that
\[ E \left[ \max_{g=1}^G \left\{ \left( \frac{N}{2} \right) \min \{ R_{s1}^g ,
   R_{s2}^g \} \right\} \right] \le E \left[ \max_{g=1}^G  \frac{NR_{s1}^g}{2}
   \right] = \left(\frac{N}{2}\right) E\left[ \log \left( 1 + \max_{g=1}^G
  |h_{\pi (\frac{N}{2}+1)}^g|^2 P \right) \right] . \]
The expression on the right is the average throughput of the best among median
users scheme proposed earlier. Using the results of Theorem~\ref{statmg}, we get
\[ R_{tot} = O ( N \log \log G). \]
\[ \Rightarrow \Omega (N) = R_{tot} = O \left( N \log \log G \right). \]
We now determine the average delay of the multi-group cooperation scheme. To
implement this scheme, the base station needs to maintain $G$ queues, one for
each group. At the beginning of each time slot, the base station selects a group
according to condition (\ref{cond5}). Since we consider a symmetric scenario,
the probability that the base station chooses any particular group is $(1/G)$.
The information transmitted by the base station in the f\mbox{}irst half of each time
slot reaches all the $N$ users in the selected group at the end of that time
slot. Hence the average delay for transmitting a packet in the multi-group
cooperation scheme is given by
\[ D = G ( T_c + \mu S) = \Theta \left( \frac{G}{E[R]} \right), \]
where $R_{tot} = (N E[R])/2$. Hence the average delay of the multi-group
cooperation scheme can be bounded as
\[ \Omega \left( \frac{G}{\log \log G} \right) = D = O (G). \]

\section{Proof of Lemma~\ref{Chiw}} \label{appChiw}

From the results on extreme order statistics in \cite{book}, we know that
\[ \frac{|\chi_{min}|^2}{b_N} \to W \quad \textrm{in distribution}, \]
where $W$ has a Weibull type distribution and $b_N$ satisf\mbox{}ies
$F(b_N) = \frac{1}{N}$, which implies
\[  1 - e^{-Lb_N} \left( \sum_{k=0}^{L-1} \frac{(Lb_N)^k}{k!} \right) =
 \frac{1}{N}. \]
Using Taylor's theorem, we get for some $0 < \gamma_N < 1$
\[ 1 -  e^{-Lb_N} \left( e^{Lb_N} - \frac{e^{\gamma_N Lb_N} (Lb_N)^L}{L!}
  \right) = \frac{1}{N} \quad \Rightarrow \frac{e^{-(1-\gamma_N) Lb_N}
  (Lb_N)^L}{L!} =  \frac{1}{N}. \]
Taking $\log(.)$ on both sides, we get
\[ (1-\gamma_N) Lb_N - L \log b_N = \log N + L \log L - \log (L!). \]
Since $|\chi_{min}|^2 \le |\chi_1|^2 = \Theta(1)$, we know that
$b_N = O(1)$ and hence the $\log b_N$ term dominates the left hand
side of the above expression. Thus we have
\[ b_N = \Theta \left( N^{-\frac{1}{L}} \right). \]
\[ \Rightarrow N^{\frac{1}{L}} |\chi_{min}|^2 \to k W  \quad
   \textrm{in distribution, for some constant $k>0$.} \]
Since $E\left[|\chi_{min}|^2 \right] \le E \left[|\chi_1|^2 \right]
< \infty$, we can use the result in Theorem~2.1 of \cite{Pick} to
conclude that \[  N^{\frac{1}{L}} E\left[|\chi_{min}|^2 \right] \to
k E[W] = \Theta(1). \]
\[ \Rightarrow  E\left[|\chi_{min}|^2 \right] = \Theta \left(
   N^{-\frac{1}{L}} \right). \]
The average throughput of the worst user scheme can now be upper
bounded using Jensen's inequality as follows
\[ R_{tot} = N E \left[ \log \left( 1 + |\chi_{min}|^2 P\right) \right]
   \le N \log \left( 1 + E \left[|\chi_{min}|^2 \right] P\right). \]
\beq \label{lbChi}
\Rightarrow R_{tot} = O \left(N^{\left( \frac{L-1}{L} \right)} \right).
\enq
We lower bound the average throughput of the worst user scheme as follows
\[ R_{tot} = N \int_{0}^{\infty} \log (1+xP) {\mathrm d} F_{min}(x) \ge N
  \int_{b_N}^{\infty} \log (1+xP) {\mathrm d} F_{min}(x) . \]
\[ \Rightarrow R_{tot} \ge N \log \left( 1+ b_N P \right)
  \left[ 1 - F_{min}(b_N) \right], \]
where
\[ F_{min}(x) = 1 - (1-F(x))^N. \]
Using the fact that $F(b_N) = \frac{1}{N}$, we get
\[ F_{min}(b_N) = 1 - (1 - F(b_N))^N = 1 - \left( 1 - \frac{1}{N}
\right)^N = 1 - e^{N \log \left( 1 - \frac{1}{N} \right)} . \]
\[ \Rightarrow F_{min}(b_N) = 1 - e^{-1} \left( 1 + O \left(
\frac{1}{N} \right) \right). \]
\[ \Rightarrow R_{tot} \ge  N \log \left( 1+ b_N P \right) \left[
e^{-1} +  O \left( \frac{1}{N} \right) \right] = \Theta \left( N
\log  \left( 1+ N^{-\frac{1}{L}} P \right) \right). \]
\[ \Rightarrow R_{tot} = \Omega \left(N^{\left( \frac{L-1}{L}
  \right)} \right). \]
Combining this with the upper bound in (\ref{lbChi}), we get
\[ R_{tot} = \Theta \left(N^{\left( \frac{L-1}{L} \right)} \right). \]

\section{Proof of Lemma~\ref{Chib}} \label{appChib}

From the results on extreme order statistics in \cite{book}, we know that
\[ \frac{|\chi_{max}|^2 - a_N}{b_N} \to W \quad \textrm{in distribution}, \]
where $W$ has a Gumbel distribution and $a_N$ and $b_N$ satisfy
\[ F(a_N) = 1 - \frac{1}{N} \quad \textrm{and} \quad b_N = \frac{1}{N f(a_N)}, \]
where $f(.)$ denotes the probability density function obtained from (\ref{chinorm}).
Now \[ F(a_N) = 1 - \frac{1}{N} \quad \Rightarrow \frac{e^{-La_N}(La_N)^{(L-1)}}
 {(L-1)!} \left( 1 + O \left(\frac{1}{a_N} \right) \right) = \frac{1}{N}. \]
Taking $\log(.)$ on both sides and simplifying, we get
\[ La_N - (L-1) \log a_N = \log N + (L-1) - \frac{1}{2} \log (L-1) + K .\]
\[ \Rightarrow a_N = \frac{\log N + (L-1) \log \log N}{L} + O(\log \log N). \]
Since \[ f(a_N) = \frac{L e^{-La_N} (L a_N)^{(L-1)}}{(L-1)!} = \Theta \left(
 \frac{1}{N} \right), \] we have $b_N = C = \Theta(1)$. Thus
\[ |\chi_{max}|^2 - \left( \frac{\log N + (L-1) \log \log N}{L} + O(\log \log N)
  \right) \to CW \quad \textrm{in distribution}. \]
Using Chebyshev's inequality, it is easy to show that
\[ \frac{|\chi_{max}|^2}{\left( \frac{\log N + (L-1) \log \log N}{L} \right)}
  \to 1 \quad \textrm{in probability}. \]
Since any Chi-squared random variable with $2L$ degrees of freedom can be
expressed as the sum of $L$ exponential i.i.d random variables, we have
\[ |\chi_{max}|^2 = \max_{i=1}^N \left\{ \frac{Z_1^i + \cdots + Z_L^i}{L}
  \right\} \le \max_{i=1}^N Z_1^i, \]
where $Z_j^i$'s are exponential random variables with unit mean. Hence
\[ E \left[ \frac{|\chi_{max}|^2}{\left(\frac{\log N + (L-1) \log \log N}{L}
 \right)} \right] \le \frac{E \left[ \max_{i=1}^N Z_1^i \right]}{\left(
 \frac{\log N + (L-1) \log \log N}{L} \right)} \le \frac{k \log N}{\left(
 \frac{\log N + (L-1) \log \log N}{L} \right)}  \le k L < \infty. \]
Thus we can apply the Dominated Convergence Theorem to get
\[  E \left[ \frac{|\chi_{max}|^2}{\left(\frac{\log N + (L-1) \log \log N}{L}
  \right)} \right] \to 1 \quad \Rightarrow E \left[ |\chi_{max}|^2 \right] =
   \Theta \left( \frac{\log N + (L-1) \log \log N}{L} \right). \]
Using Jensen's inequality, we get
\[ R_{tot} = E \left[ \log \left( 1 + |\chi_{max}|^2 P\right) \right] \le
   \log  \left( 1 + E \left[ |\chi_{max}|^2 \right] P \right) .\]
\beq \label{ubChi}
\Rightarrow R_{tot} = O \left( \log  \left( 1 +  \frac{\log N +
 (L-1) \log \log N}{L} \right) \right).
\enq
We can lower bound the average throughput of the best user scheme as follows
\[ R_{tot} = \int_{0}^{\infty} \log (1+xP) {\mathrm d} F_{max}(x) \ge
  \int_{a_N}^{\infty} \log (1+xP) {\mathrm d} F_{max}(x) . \]
\[ \Rightarrow R_{tot} \ge \log \left( 1+ a_N P \right)
  \left[ 1 - F_{max}(a_N) \right], \]
where
\[ F_{max}(x) = (F(x))^N. \]
Using the fact that $F(a_N) = 1 - \frac{1}{N}$, we get
\[ F_{max}(a_N) = (F(a_N))^N = \left( 1 - \frac{1}{N} \right)^N =
 e^{-1} \left( 1 + O \left( \frac{1}{N} \right) \right). \]
\[ \Rightarrow R_{tot} \ge \log \left( 1+ a_N P \right) \left[
1 - e^{-1} +  O \left( \frac{1}{N} \right) \right] = \Theta \left(
\log  \left( 1+ a_N P \right) \right). \]
\[ \Rightarrow R_{tot} = \Omega \left( \log \left( 1 +  \frac{\log N +
 (L-1) \log \log N}{L} \right) \right). \]
Combining this with the upper bound in (\ref{ubChi}), we get
\[ R_{tot} = \Theta \left( \log \left( 1 +  \frac{\log N +
 (L-1) \log \log N}{L} \right) \right). \]

\newpage
\bibliographystyle{ieeetr}
\bibliography{draft}

\begin{thebibliography}{10}

\bibitem{berry}
R.~Berry and R.~Gallager, ``Communication over fading channels with delay
  constraints,'' {\em IEEE Transactions on Information Theory}, vol.~48,
  pp.~1135--1149, May 2002.

\bibitem{ashu}
D.~Rajan, A.~Sabharwal, and B.~Aazhang, ``Delay bounded packet scheduling of
  bursty sources over wireless channels,'' {\em IEEE Transactions on
  Information Theory}, vol.~50, pp.~125--144, January 2004.

\bibitem{yeh}
E.~M. Yeh and A.~S. Cohen, ``Information theory, queueing, and resource
  allocation in multi-user fading communications,'' in {\em 38th Annual
  Conference on Information Sciences and Systems}, March 2004.

\bibitem{zhang}
J.~Zhang and D.~Zheng, ``Ad hoc networking over fading channels: Multi-channel
  diversity, mimo signaling, and opportunistic medium access control,'' in {\em
  41st Allerton Conference on Communications, Control, and Computing}, October
  2003.

\bibitem{berry2}
P.~Liu, R.~Berry, and M.~Honig, ``Delay-sensitive packet scheduling in wireless
  networks,'' in {\em IEEE WCNC 2003}, March 2003.

\bibitem{sanjay}
S.~Shakkottai and A.~Stolyar, ``Scheduling for multiple flows sharing a
  time-varying channel: The exponential rule,'' {\em American Mathematical
  Society Translations, Series 2}, vol.~207, 2002.

\bibitem{Sasw}
P.~Chaporkar and S.~Sarkar, ``On-line optimal wireless multicast,'' in {\em 2nd
  Workshop On Modeling and Optimization in Mobile, Ad Hoc and Wireless
  Networks}, (Cambridge, England), pp.~282--291, March 2004.

\bibitem{Tay}
C.~Wu and Y.~Tay, ``Amris: A multicast protocol for ad hoc wireless networks,''
  in {\em IEEE MILCOM'99}, (Atlantic City, NJ), November 1999.

\bibitem{Aceves}
J.~Garcia-Luna-Aceves and E.~Madruga, ``The core-assisted mesh protocol,'' {\em
  IEEE Journal on Selected Areas in Communications}, vol.~17, pp.~1380--1394,
  August 1999.

\bibitem{Knopp}
R.~Knopp and P.~Humblet, ``Information capacity and power control in single
  cell multiuser communications,'' in {\em IEEE International Computer
  Conference (ICC'95)}, (Seattle, WA), June 1995.

\bibitem{erkip}
A.~Sendonaris, E.~Erkip, and B.~Aazhang, ``User cooperation diversity-part i:
  System description,'' {\em IEEE Transactions on Communications}, vol.~51,
  pp.~1927--1938, November 2003.

\bibitem{Prav}
P.~K. Gopala and H.~E. Gamal, ``Opportunistic multicasting,'' in {\em Asilomar
  Conference on Signals, Systems and Computers}, November 2004.

\bibitem{cover}
T.~Cover and J.~Thomas, {\em Elements of Information Theory}.
\newblock New York: John Wiley Sons, Inc., 1991.

\bibitem{Hass1}
M.~Sharif and B.~Hassibi, ``Delay analysis of throughput optimal scheduling in
  broadcast fading channels,'' {\em Submitted to IEEE Transactions on
  Information Theory}, 2004.

\bibitem{Newman}
D.~J. Newman and L.~Shepp, ``The double dixie cup problem,'' {\em Amer. Math.
  Monthly}, vol.~67, pp.~58--61, January 1960.

\bibitem{Feller}
W.~Feller, {\em An introduction to probability theory and its applications}.
\newblock John Wiley and Sons, Inc., 1967.

\bibitem{Johnson}
N.~L. Johnson and S.~Kotz, {\em Urn models and their application}.
\newblock John Wiley and Sons, Inc., 1977.

\bibitem{Caire}
G.~Caire and D.~Tuninetti, ``The throughput of hybrid-arq protocols for the
  gaussian collision channel,'' {\em IEEE Transactions on Information Theory},
  vol.~47, July 2001.

\bibitem{Hass2}
M.~Sharif and B.~Hassibi, ``On the capacity of mimo broadcast channel with
  partial side information,'' {\em To appear in IEEE Transactions on
  Information Theory}, 2004.

\bibitem{book}
B.~C. Arnold, N.~Balakrishnan, and H.~N. Nagaraja, {\em A first course in order
  statistics}.
\newblock New York: John Wiley Sons, Inc., 1992.

\bibitem{Durrett}
R.~Durrett, {\em Probability: Theory and Examples}.
\newblock California: Duxbury Press, Inc., 1996.

\bibitem{intbook}
I.~S. Gradshteyn and I.~M. Ryzhik, {\em Table of Integrals, Series, and
  Products}.
\newblock Academic Press, Inc., 1980.

\bibitem{Pick}
J.~Pickands, ``Moment convergence of sample extremes,'' {\em The Annals of
  Mathematical Statistics}, vol.~39, no.~3, pp.~881--889, 1968.

\end{thebibliography}

\newpage

\newpage

\newpage

\newpage

\newpage

\newpage

\end{document}